\documentclass{emulateapj}
\usepackage{epsfig}
\usepackage{natbib}
\usepackage{color}
\usepackage{soul}
\usepackage{graphicx}
\usepackage{epstopdf}
\newcommand{\magB}{\rm \textit{\textbf{B}}}
\newcommand{\velv}{\rm \textit{\textbf{v}}}

\newcommand{\be}{\begin{eqnarray}}
\newcommand{\ee}{\end{eqnarray}}
\newcommand{\beq}{\begin{equation}}
\newcommand{\eeq}{\end{equation}}
\renewcommand{\vec}[1]{\mbox{\boldmath $\displaystyle #1$}}
\newcommand{\grad}{{\mbox{\boldmath $\nabla$}}}

\begin{document}
\title{Magnetohydrodynamic Simulations of Hot Jupiter Upper Atmospheres}
\author{George B. Trammell, Zhi-Yun Li \& Phil Arras }
\affil{Department of Astronomy, University of Virginia,
P.O. Box 400325, Charlottesville, VA 22904-4325}
\email{ gbt8f@virginia.edu, zl4h@virginia.edu,arras@virginia.edu }
\slugcomment{\ }

\begin{abstract}
Two-dimensional simulations of hot Jupiter upper atmospheres including the
planet's magnetic field are presented. The goal is to explore magnetic effects
on the layer of the atmosphere that is ionized and heated by stellar EUV
radiation, and the imprint of these effects on the Ly$\alpha$ transmission
spectrum. The simulations are axisymmetric, isothermal, and include both
rotation and azimuth-averaged stellar tides. Mass density is converted to atomic
hydrogen density through the assumption of ionization equilibrium. The
three-zone structure -- polar dead zone, mid-latitude wind zone, and equatorial
dead zone -- found in previous analytic calculations is confirmed. For a
magnetic field comparable to that of Jupiter, the equatorial dead zone, which is
confined by the magnetic field and corotates with the planet, contributes at
least half of the transit signal. For even stronger fields, the gas escaping in
the mid-latitude wind zone is found to have a smaller contribution to the
transit depth than the equatorial dead zone. Transmission spectra computed from
the simulations are compared to HST STIS and ACS data for HD 209458b and HD
189733b, and the range of model parameters consistent with the data is found.
The central result of this paper is that the transit depth increases strongly
with magnetic field strength when the hydrogen ionization layer is magnetically
dominated, for dipole magnetic field $B_0 \ga 10\ {\rm G}$. Hence transit depth
is sensitive to magnetic field strength, in addition to standard quantities such
as the ratio of thermal to gravitational binding energies. Another effect of the
magnetic field is that the planet loses angular momentum orders of magnitude
faster than in the non-magnetic case, because the magnetic field greatly
increases the lever arm for wind braking of the planet's rotation. Spin-down
timescales for magnetized models of HD 209458b that agree with the observed
transit depth can be as short as $\simeq 30\ {\rm Myr}$, much shorter than the
age of the system.

\end{abstract}

\section{Introduction}

Hot Jupiters are gas giants orbiting close to their parent stars.
The large stellar EUV flux heats and ionizes the upper atmosphere of these
planets, increasing the thermal energy to a value approaching the 
gravitational binding energy, leading to a region  weakly bound
to the planet. The resulting large gas scale heights and atmospheric escape form
an extended upper atmosphere around the planet, which may be probed by 
transmission spectroscopy using strong atomic resonance lines.

The existence of an extended upper atmosphere has been established through a
variety of observations. Spectroscopic UV observations of HD 209458b
(\citealt{Henry 2000}) indicate a $\sim 10\%$ decrease in flux during transit at
$\sim 100\ \rm km\ s^{-1}$ from the center of the hydrogen Ly$\alpha$ line. This
transit depth has been attributed to an atmosphere of neutral H extending to a
radius $\simeq 2.4 R_p$, where $R_p$ is the radius of the broadband
photosphere of the planet (see \citealt{Vidal 2008}). As this radius is
comparable to the Roche lobe radius, \citet{Vidal 2004} suggested that the
planet is losing mass through Roche lobe overflow.

Additional observations of HD 209458b at transit have indicated absorption
in other resonance lines, including NaI \citep{Charbonneau 2002,
Sing 2008}, OI \citep{Vidal 2004}, CII \citep{Vidal 2004, France 2011}, 
and SiIII \citep{Linsky 2010, France 2011}. Follow-up observations
and re-analysis of HST-ACS data, in comparison with HST-STIS low and
medium-resolution spectra, confirmed the reduction of Ly$\alpha$
flux \citep{Ehrenreich 2008}.

Transmission spectra of the hot Jupiter HD 189733b have also revealed
absorption due to HI (in both Ly$\alpha$ and H$\alpha$;
\citealt{Lecavelier 2010} and \citealt{2012ApJ...751...86J}) and 
NaI \citep{Redfield2008, Snellen 2008}. HST-COS observations by \citet{Linsky 2010}
have indicated absorption at up to $\pm50$ km s$^{-1}$ from line center
in CII and SiIII that may be indicative of high velocity absorbers in
the upper atmosphere (although these observations probe deeper layers
than those probed by the HI observations).  Multi-epoch spectra have
also revealed significant changes in the Ly$\alpha$ transit depth,
which are correlated with flares in ionizing radiation from the
host star detected with HST and SWIFT \citep{Lecavelier 2012}.

This paper will focus on the Ly$\alpha$ absorption observed
in the upper atmospheres of HD 209458b and HD 189733b. One 
interpretation of this absorption invokes hydrogen with thermal velocity
$\sim 10\ \rm km\ s^{-1}$ with such a large column density that the damping
wings of Ly$\alpha$ become optically thick (e.g. Yelle 2004). 
Alternatively, a much smaller column is required if 
hydrogen atoms at thermal velocities $\sim 100\ 
\rm km\ s^{-1}$, created by charge exchange with stellar wind protons
\citep{2008Natur.451..970H, 2010ApJ...709..670E, Tremblin 2012},
produce a sufficiently broad line profile.

This paper considers the former scenario in which the transit
depth is due to a layer of thermal hydrogen in the planet's
atmosphere. A number of studies have already explored the properties 
of strongly irradiated exoplanet atmospheres, and the possibility 
of thermally-driven hydrodynamic outflow \citep{Yelle
2004, Yelle 2006, 2005ApJ...621.1049T, Garcia 2007, Murray-Clay
2009, Ehrenreich 2011} and/or Roche Lobe overflow 
\citep{2003ApJ...588..509G, 2010Natur.463.1054L, 2010ApJ...721..923L, 
Ehrenreich 2011}. The present study stands apart from the previous 
ones by including, through detailed magnetohydrodynamic (MHD) 
simulations, the effect of the planetary magnetic field. It is a 
follow-up of Trammell et al. (2011), which considered the 
magnetic effects semi-analytically.

\citet[][hereafter, Paper I]{Trammell 2011} showed that the addition of the
planetary magnetic field leads to the formation of an equatorial ``dead-zone"
(DZ) --- a static region where the wind ram pressure is insufficient to
overwhelm magnetic stresses and open the field lines into an outflow. This
effect is well known in the classical MHD stellar wind theory (e.g. Mestel
1968). Paper I found a second static region near the poles where the wind can be
shut off by the increased gravitational potential barrier from the stellar tide.
In the strong tide limit, a wind-zone (i.e., the outflow region; WZ) is then
expected to exist only at intermediate latitudes. A goal of the present paper is
to verify this analytically-obtained three-zone structure with detailed
numerical simulations.

Another conclusion from Paper I was that observations of Ly$\alpha$ absorption
at the 5-10\% level for HD 209458b may be detecting neutral H which is
collisionally coupled to ionized gas confined to the equatorial DZ by the
planet's magnetic field. The bulk of the absorbing gas observed at transit thus
may not be escaping, but rather is in the static equatorial dead zone. This
qualitative result differs from the basic assumption in the hydrodynamic escape
and Roche Lobe overflow models, that the transit observations are probing gas in
the act of escaping from the planet.
 
A limitation of the analytic models in Paper I is that they ignore magnetic
forces, which means that the poloidal magnetic geometry was assumed rather than
computed self-consistently. Another limitation is that the fluid was assumed to
corotate with the planet everywhere, including the wind zone, where the
corotation is expected to break down at large distances. The treatment in this
paper overcomes these limitations by performing MHD simulations, which compute
the magnetic field structure and fluid rotation self-consistently. This allows a
more accurate calculation of the mass and angular momentum loss rates, as well
as the density and velocity profiles required to compute transmission spectra.

The plan of the paper is as follows.
Section \ref{zeus.sec} describes the simulation setup and model
parameters, and Section \ref{sims.sec} presents the simulation results.
Section \ref{transit.sec} describes the method for computing
model Ly$\alpha$ spectra. The frequency-dependent and
frequency-integrated transit depths for a range of simulation parameters
are compared with observations of HD 209458b and HD 189733b.
Findings are summarized in Section \ref{summary.sec}. The Appendix 
contains a discussion of numerical effects at the shear layer separating the
dead and wind zones.

\section{Simulation Setup}
\label{zeus.sec}

Consider a planet of mass $M_p$ and radius $R_p$ in a circular orbit at a distance $D$ from a star of
mass $M_\star$. The planet's rotation is synchronized to the orbit with angular velocity
$\Omega_p=[G(M_\star+M_p)/D^3]^{1/2}$, and the spin axis is aligned with the orbital 
angular momentum. Outside the planet, in the region modeled by the simulations, the gas 
is not required to corotate with the planet.

Two-dimensional (2D), axisymmetric simulations in spherical coordinates
are carried out with the publicly available MHD code
ZEUS-MP \citep[see][and references therein]{Hayes 2006}, which solves
the ideal-MHD equations: 
\begin{eqnarray}
{\frac{D \rho}{D t}} &=& {-\rho \nabla \cdot \velv \label{cont.eq}} \\
{\rho \frac{D \velv}{D t}}   &=& {- \nabla P + \frac{1}{4 \pi}
                                (\nabla \times \magB) \times \magB - \rho \nabla U 
                                \label{pressure.eq}} \\
{\frac{\partial \magB}{\partial t}} &=& {\nabla \times (\velv \times \magB)
                                  \label{curlB.eq},}
\end{eqnarray}
where the comoving (Lagrangian) derivative is defined as
\begin{eqnarray}
\frac{D}{D t} &\equiv& \frac{\partial}{\partial t} + 
                       {\velv \cdot \nabla} \label{deriv.eq}.
\end{eqnarray}
Equations \ref{cont.eq}-\ref{curlB.eq} are the mass continuity, 
momentum and induction equation, respectively. The neglect of explicit fluid viscosity in Equation \ref{pressure.eq} and resistivity in Equation \ref{curlB.eq} are discussed in Appendix B of Paper I. The quantity 
$U$ is an effective potential to be defined below, and other symbols
have their standard meaning. 

Instead of solving the energy equation, an isothermal equation of state, $P=\rho
a^2$, is used, where $a$ is the (constant) isothermal sound speed. This
assumption is equivalent to adding energy to the flow to counter adiabatic
cooling. It gives rise to a transonic outflow (e.g. \citealt{lc99}). The
isothermal assumption is convenient for the present study, where the focus is
not on the initial launching of the wind, but rather on magnetic effects. A more
detailed study, beyond the scope of this paper, would include heating and
cooling effects in an energy equation. Note, however, that since the magnetic
field and rotation are included, the flow is also accelerated in part by the
``magneto-centrifugal" effect \citep{1982MNRAS.199..883B}, as well as stellar
tides.

The computational grid extends from an inner radial boundary at 
$r=R_p$ to the outer boundary at $r=30 R_p$, with $R_p$ set to the planet's observed 
transit continuum radius \citep{Southworth 2010}, and from the north
pole at $\theta=0$ to the south pole at $\theta=\pi$. The radial box 
size was chosen through experimentation so that all MHD critical points 
in the wind 
zone were contained within the computational domain for a wide 
range of model parameters. The standard resolution is $272\times 200$,
with the radial cell size $\Delta r = r_{i+1}-r_i$ increasing outward
according to $\Delta r_{i+1}/\Delta r_i = 1.02$; the ratio was chosen 
to adequately resolve the wind acceleration 
region near the base. The $\theta$ grid is uniformly spaced. Surrounding
the active grid are two layers of ghost zones at each of the four 
boundaries; they are used to impose boundary conditions. 


At time $t=0$, the fluid is uniformly rotating with velocity
$(v_r,v_\theta,v_\phi)=(0,0,\Omega_p r \sin\theta)$, and in hydrostatic balance
over most of the computational grid. The initial magnetic field is assumed to be a potential
field, so that the magnetic force is everywhere zero, an assumption consistent with the equation of hydrostatic balance. In a reference frame corotating with the
planet, and with the origin comoving with the planet, hydrostatic balance takes the
form (Paper I)
\be
0 & = & - a^2 \grad \rho - \rho \grad U_{\rm rot}.
\label{eq:hb}
\ee
The potential $U_{\rm rot}$ includes contributions 
from the gravity of the planet, the stellar gravity, the dipole term arising from the acceleration of the origin, and the centrifugal  force, and takes the form
\be 
U_{\rm rot}(\vec{x})  &=& - \frac{GM_p}{|\vec{x}|} -
\frac{GM_\star}{|\vec{x}-\vec{x}_\star|} \nonumber \\
&+& \frac{GM_\star \vec{x} \cdot \vec{x}_\star}{|\vec{x}_\star|^3}
- \frac{1}{2} \left| \vec{\Omega_p}\times \vec{x} \right|^2
\label{eq:Urotgen}
\ee
The dipole term, which acts to accelerate the center of mass of the planet, cancels off part of the stellar gravity, leaving only a tidal acceleration. Expressing the position vector in spherical coordinates $\vec{x}=(r,\theta,\phi)$, the position of the star as $\vec{x}_\star=(D,\pi/2,0) = D\vec{e}_x$, and making the tidal approximation, $r \ll D$, gives
\be 
U_{\rm rot}(\vec{x})  &\simeq& -  \frac{GM_p}{r} - \frac{1}{2} \Omega_p^2 r^2
\left( f_{\rm rot} \sin^2\theta - 1 \right),
\label{eq:Urot}
\ee
where the longitude-dependent function $f_{\rm rot}=1+3\cos^2\phi$.
As the simulations are axisymmetric, we substitute the azimuthal average
$\cos^2\phi \rightarrow 1/2$, yielding $f_{\rm rot}=5/2$. At the equator, the radial acceleration $\partial U_{\rm rot}/\partial r = 0$ at the radius
\be
r_{\rm H} & = & \left( \frac{2 GM_p}{3\Omega_p^2} \right)^{1/3} \simeq D \left( \frac{2M_p}{3M_\star}\right)^{1/3},
\label{eq:rl}
\ee
a factor of $2^{1/3}$ larger than the physically correct value for the Lagrange points, which are evaluated along the star-planet line. In this paper $r_{\rm H}$ will be called the Hill radius.

Substituting Equation \ref{eq:Urot} into Equation \ref{eq:hb}, the initial density distribution over the inner part of the computational domain takes the form
\begin{eqnarray}
\rho(r,\theta) &=& \rho_{\rm ss} \ {\rm exp}\left[
- \left( \frac{U_{\rm rot}(r,\theta) -U_{\rm rot}(R_p,\pi/2)}{a^2} \right) \right],
\label{density}
\end{eqnarray}
where $\rho_{\rm ss}$ is the density at $(r,\theta)=(R_p,\pi/2)$. 
The initial density distribution then requires $a$ and $\rho_{\rm ss}$ as parameters, in addition to the parameters of the planet, star and orbit. Well outside $r_{\rm H}$, the hydrostatic density profile rises steeply to large values, due to the net acceleration pointing outward. For the initial condition only, a ceiling is placed on the density at $\rho \leq \rho_{\rm ss}$ to limit this growth. In the outer regions of the computational grid where $\rho=\rho_{\rm ss}$, the pressure force is then zero and the tidal and centrifugal forces pull mass outward, initiating the outflow. Since the physical, steady-state solutions exhibit $\rho \ll \rho_{\rm ss}$ in the outer regions, capping the density is a device to allow the density to decrease to physical levels more quickly.

It is simpler to perform the simulation not in a corotating frame, which would require the addition of Coriolis and centrifugal forces, but rather in a non-rotating frame. The origin still moves with the center of the planet. In this reference frame, the centrifugal term can be omitted from the potential, giving 
\be 
U(\vec{x})  &=& - \frac{GM_p}{|\vec{x}|} - \frac{GM_\star}{|\vec{x}-\vec{x}_\star|}
+ \frac{GM_\star \vec{x} \cdot \vec{x}_\star}{|\vec{x}_\star|^3}
\nonumber \\
& \simeq & 
-  \frac{GM_p}{r} - \frac{1}{2} \Omega_p^2 r^2
\left( f \sin^2\theta - 1 \right),
\label{eq:U}
\ee
where now $f=1+\cos^2\phi$. The azimuthal average gives $f=3/2$ for the non-rotating frame. The $r$ and $\theta$ components of $-\grad U$ from Equation \ref{eq:U} are introduced into the Zeus-MP code as a source term in the momentum equations. Since the gas is not required to corotate, Equation \ref{eq:rl} may underestimate the radius at which the equatorial acceleration changes sign. An upper limit is found by ignoring the centrifugal force. Using $f=3/2$ with Equation \ref{eq:U} would give $(2GM_p/\Omega_p^2)^{1/3}$ for this radius, larger than the expression in Equation \ref{eq:rl} by a factor $3^{1/3}$.

The initial condition for the magnetic field is a dipole with magnetic axis
aligned with the rotation axis:
\begin{eqnarray}
B_r &=& B_0 \left(\frac{r}{R_p}\right)^{-3} \cos \theta \\
B_{\theta} &=& \frac{B_0}{2} \left(\frac{r}{R_p}\right)^{-3} \sin \theta \\
B_\phi & = & 0,
\end{eqnarray}
where the field at the magnetic pole is $B_0$. The development of nonzero
$B_\phi$ at $t > 0$ will lead to magnetic torques on the gas and planet. A key parameter of the model is
the equatorial value of the plasma $\beta$ at the inner radius:
 \be
 \beta_0 & = & \frac{ 8\pi  P_{\rm ss}}{(B_0/2)^2}
 \ee
 where $P_{\rm ss}=a^2 \rho_{\rm ss}$ is the base pressure at the equator, and
 $B_0/2$ is the magnetic field at the equator. This parameter sets the size of the equatorial DZ (Paper I). 
 
Compared with the initial conditions, the boundary conditions are much 
more difficult to implement, especially on the inner radial boundary, which is the base of both the wind and dead zones. Since 
the boundary conditions are crucial to the success of the numerical 
simulations, they will be described in some detail. 


Consider first the boundary condition on
$\rho$ at the inner radial boundary. The densities in 
the inner radial ghost zones {\it and the first radial 
active zones} are kept at their initial values at all times. Even 
though the densities in the first active zones are updated at each 
time step, the updated values are discarded and replaced by their 
initial values. This guarantees that the base density is held fixed 
at the prescribed value, even in the wind zone. 

For the inner radial boundary, $v_r=0$ is set
at the inner face of the first active zone, as well as in the ghost zones.
In other words,  $v_r(R_p,\theta)=0$ and $v_r(r^-,\theta)$=0, where 
$r^-$ ($< R_p$) denotes the inner radial ghost region.
The reflection boundary condition is applied to $v_\theta$, so that
$v_\theta(r^-,\theta)=v_\theta(r^+,\theta)$, where $r^+$ is the 
symmetry point (with respect to the $r=R_p$ surface) in the 
active domain of the location $r^-$ in the ghost region. The
boundary condition on the 
azimuthal velocity component, $v_\phi$, is $v_\phi(r^-,\theta) 
= 2 v_\phi(R_p,\theta)-v_\phi(r^+,\theta)$ where
$v_\phi(R_p,\theta)=\Omega_{\rm p} R_p \sin \theta$. That is, the average
of the first ghost zone and the first active zone should equal
the corotation velocity. It has been
verified that, in the absence of magnetic field, rotation and 
stellar tides, the inner hydro boundary conditions produce a 
thermally driven wind that matches, in steady state, the 
well-known analytic solution. 


The magnetic boundary conditions at the inner radial boundary are 
more complicated to implement. They are enforced through the 
electromotive force (EMF) ${\vec \epsilon} =\velv \times {\vec B}$, 
as this will automatically preserve $\partial(\nabla\cdot {\vec B})/\partial t =0$ during the
time evolution. 
In 2D (axisymmetric) geometry, only the $r$- and
$\theta$-components of $\vec \epsilon$ affect $B_\phi$:
\be
{\partial B_\phi\over \partial t} = {1\over r}\left[{\partial\over
      \partial r} (r\epsilon_\theta) -{\partial
      \epsilon_r\over\partial \theta}\right].
\ee
Although $B_\phi$ is assumed to be zero initially in our simulation,
it can grow with time, particularly in the outflow region. The 
boundary conditions on $\epsilon_r$ and $\epsilon_\theta$ are designed
to enable $B_\phi$ in the ghost zones to grow at the same rate as in 
the active zones. Specifically, we demand 
\be
{\epsilon_r(r^-)\over r^-}={\epsilon_r(r^+)\over r^+},
\ee
and 
\be
\left[{1\over r}{\partial \over \partial
    r}(r\epsilon_\theta)\right]_{r^-}
=\left[{1\over r}{\partial \over \partial r}(r\epsilon_\theta)\right]_{r^+}.
\label{etheta}
\ee
The value of $\epsilon_\theta$ in the ghost zone is determined by equation~\ref{etheta}
together with the condition
\be
\epsilon_\theta(R_p,\theta)=\Omega_p R_p \sin\theta B_r(R_p,\theta), 
\ee
which ensures that the footpoints of the magnetic field lines corotate 
with the planet. 

For $\epsilon_\phi=v_r B_\theta-v_\theta B_r$, the boundary condition
$\epsilon_\phi(R_p,\theta)=0$ enforces poloidal velocity parallel to poloidal
magnetic field at $r=R_p$. The use of this ``flux freezing" condition for
the magnetic field is justified in Appendix B of Paper I. It also guarantees
that $B_r(r=R_p)$ remains unchanged, i.e., the footpoints of the field lines are
firmly anchored on the rotating inner radial boundary. For $\epsilon_\phi$ in
the ghost zone, $\epsilon_\phi(r^-)=-\epsilon_\phi(r^+)$ is enforced so that the
radial gradient of $\epsilon_\phi$, which controls the evolution of $B_\theta$,
is continuous across the inner radial boundary. This set of magnetic boundary
conditions is similar to that used successfully by Krasnopolsky et al. (1999,
2003) to simulate disk-driven magnetocentrifugal winds.

%
%

The standard ``outflow" boundary condition implemented in ZeusMP is used at the outer radial boundary, with all hydrodynamic 
variables and the three components of the EMF projected to zero 
slope. In addition, $\partial B_{\phi}/\partial t=0$ is set for 
the outer radial ghost zones, which was found to prevent the growth 
of unphysically large currents that sometimes develop near the outer 
boundary. At the $\theta=0$ and $\theta=\pi$ boundaries, the 
standard ``axial'' boundary condition as implemented in ZeusMP is used, which 
enforces reflection symmetry for the $r$-component of the velocity and
magnetic field; the $\theta$- and $\phi$-components are reflected 
with a change of sign. 

%
%



The simulations were evolved until a steady-state solution was achieved.
A summary of the main model parameters
is shown in Tables \ref{runs.tab} and \ref{summary.tab}. 

\begin{deluxetable}{lcc}
\tabletypesize{\small}
\tablewidth{0pt}
\tablecolumns{3}
\tablecaption{Primary Model Parameters}
\tablehead{
\colhead{Parameter} &
\colhead{Range} &
\colhead{Description}
}
\startdata
$R_p$ & 1.35 $R_{\rm Jup}$                  & planet radius \\
$\rho_{\rm ss}$ & $\sim 10^{-16}-10^{-13}$ g cm$^{-3}$ & substellar point mass density \\
$a$ & 9-11 km s$^{-1}$                   & isothermal sound speed \\
$B_0$ & 1.0-100 G                        & polar magnetic field strength \\
$M_p$ & 0.7 $M_{\rm Jup}$                  & planet mass \\
$M_\star$ & 1.1 $M_{\sun}$            & host stellar mass \\
$D$ & 0.035-0.06 AU                         & orbital separation \\
\hline
\enddata
\tablecomments{Description and range of the model parameters used in the simulations.}
\label{runs.tab}
\end{deluxetable}
\begin{deluxetable}{cccccc}
\tablewidth{0pt}
\tablecolumns{6}
\tablecaption{Parameters for HD 209458b Runs}
\tablehead{
\colhead{Run} &
\colhead{$D$ (AU)} &
\colhead{$P_{\rm ss}$ ($\mu$bar)} &
\colhead{$a$ (km/s)} &
\colhead{$B_0$ (G)} &
\colhead{$\beta_0$}
}
\startdata
Model 1         & 0.047 & 0.05 & 10.0 & 10.0 & 0.051 \\  
Model 2         & 0.047 & 0.05 & 10.0 & 1.0 & 5.1    \\  
Model 3         & 0.047 & 0.05 & 10.0 & 50.0 & 0.002    \\  
Model 4         & 0.047 & 0.05 & 9.0  & 10.0 & 0.041     \\ 
Model 5         & 0.047 & 0.05 & 11.0 & 10.0 & 0.061     \\ 
Model 6         & 0.047 & 0.005 & 10.0 & 10.0 & 0.0051   \\ 
Model 7         & 0.047 & 0.5   & 10.0 & 10.0 & 0.51     \\ 
Model 8         & 0.035 & 0.05 & 10.0 & 10.0 & 0.051     \\  
Model 9         & 0.06  & 0.05 & 10.0 & 10.0 & 0.051   \\  
Model 10	& 0.047 & 0.05 & 10.0 & 100.0 & 0.0005 
\enddata
\tablecomments{ This table contains simulation parameters varying a single parameter ($B_0, D, a, P_{\rm ss}$)
relative to the fiducial case (Model 1).
The planetary radius and mass are fixed to $R_p = 1.35\ R_J, M_p = 0.7\ M_J$.
The values of $\beta_0$ can be compared to the value $\beta_{\rm 0,Jup} = 0.069$
using Jupiter's magnetic field
and a base pressure $P_{\rm ss} = 0.05\ \mu$bar.
}
\label{summary.tab}
\end{deluxetable}

\begin{figure*}[ht]
\epsscale{1.15}
\plotone{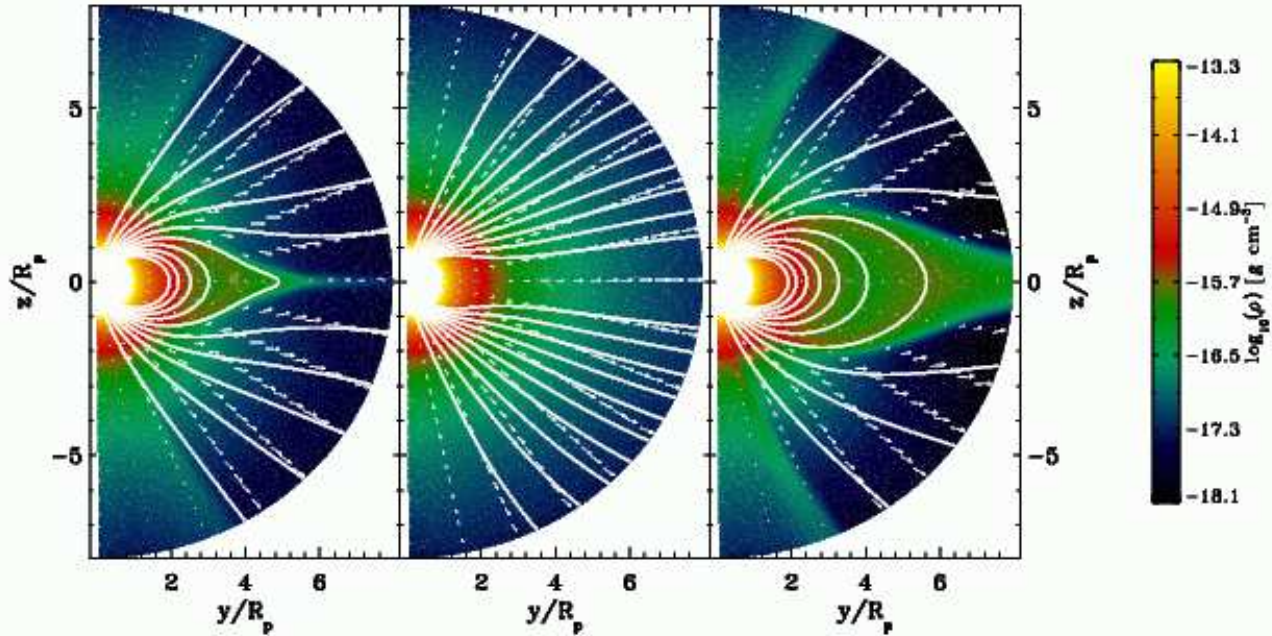}
\vspace{-8pt}
\caption{ 
Contours of total gas density $\rho(r,\theta)$ in units of g cm$^{-3}$
as viewed during transit, illustrating the effect of the magnetic
field (white lines) on the size of the equatorial DZ. {\it Left}:
Model 1 (fiducial model with $B_0 = 10\ $G) {\it Center}: Model 2
($B_0 = 1\ $G) {\it Right}: Model 3 ($B_0 = 50\ $G). For these
three models, the remaining parameters given in Table \ref{summary.tab}
are otherwise identical.  White
arrows indicate the direction and magnitude of the poloidal fluid velocity.
Only the inner 8 $R_p$ portion of the grid is shown (the
magnetic field lines  near the planet are not drawn for clarity).}
\label{fiducial_rho.fig}
\end{figure*}

\begin{figure*}[ht]
\epsscale{1.15}
\plotone{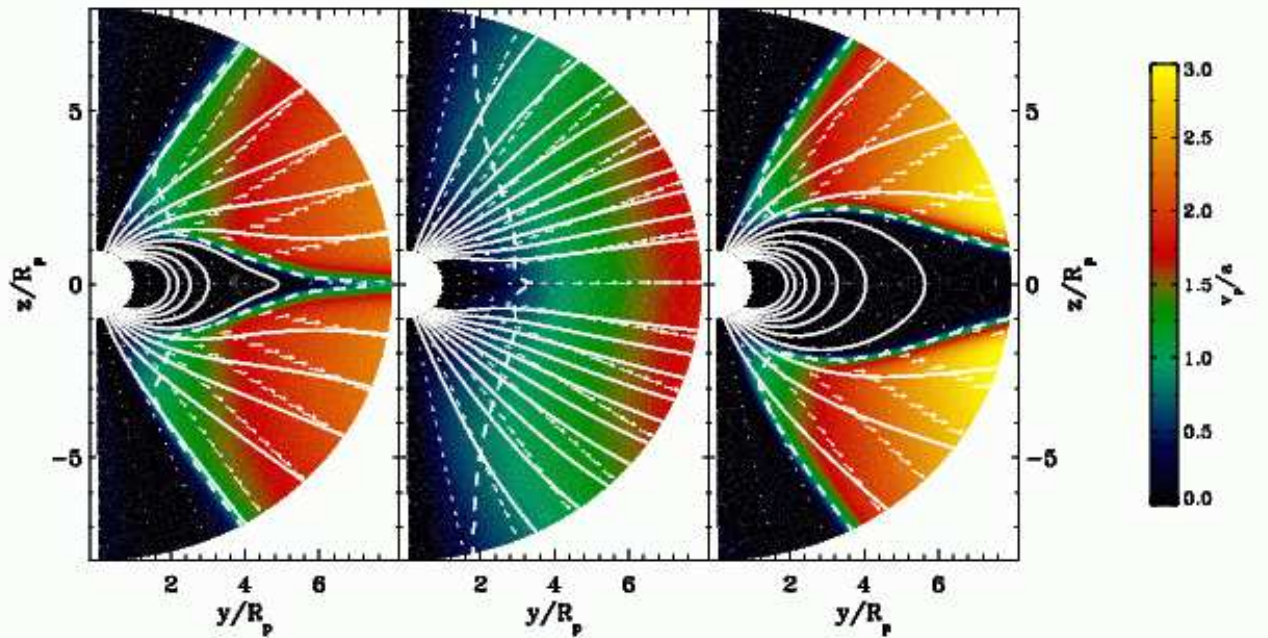}
\vspace{-8pt}
\caption{ 
Similar to Figure \ref{fiducial_rho.fig}, but for contours of poloidal
velocity $v_p$ in units of the isothermal
sound speed $a$. {\it Left}: Model 1 (fiducial model)
 {\it Center}: Model 2 {\it Right}: Model 3 for parameters given in
 Table \ref{summary.tab}. White-dashed contours trace the (slow magneto-)sonic points, which trace the vicinity of the shear layers separating the static DZs (i.e., darkest contours) from the transonic WZs.
}
\label{fiducial_v1.fig}
\end{figure*}

\section{Simulation Results}
\label{sims.sec}

A fiducial model is chosen with HD 209458b's parameters
and $B_0=10\ $G (Model 1 in Table \ref{summary.tab}). 
The other simulations listed in Table \ref{summary.tab} vary the 
model parameters listed in Table \ref{runs.tab}. A 
qualitative discussion of the simulation results is given in
\S~\ref{sec:qualitative}, and a more quantitative
analysis in \S~\ref{quantitative}. The last 
subsection (\S~\ref{rates}) contains a discussion of mass 
and angular momentum loss rates from the planet. 

\begin{figure*}[t]
\epsscale{1.15}
\plotone{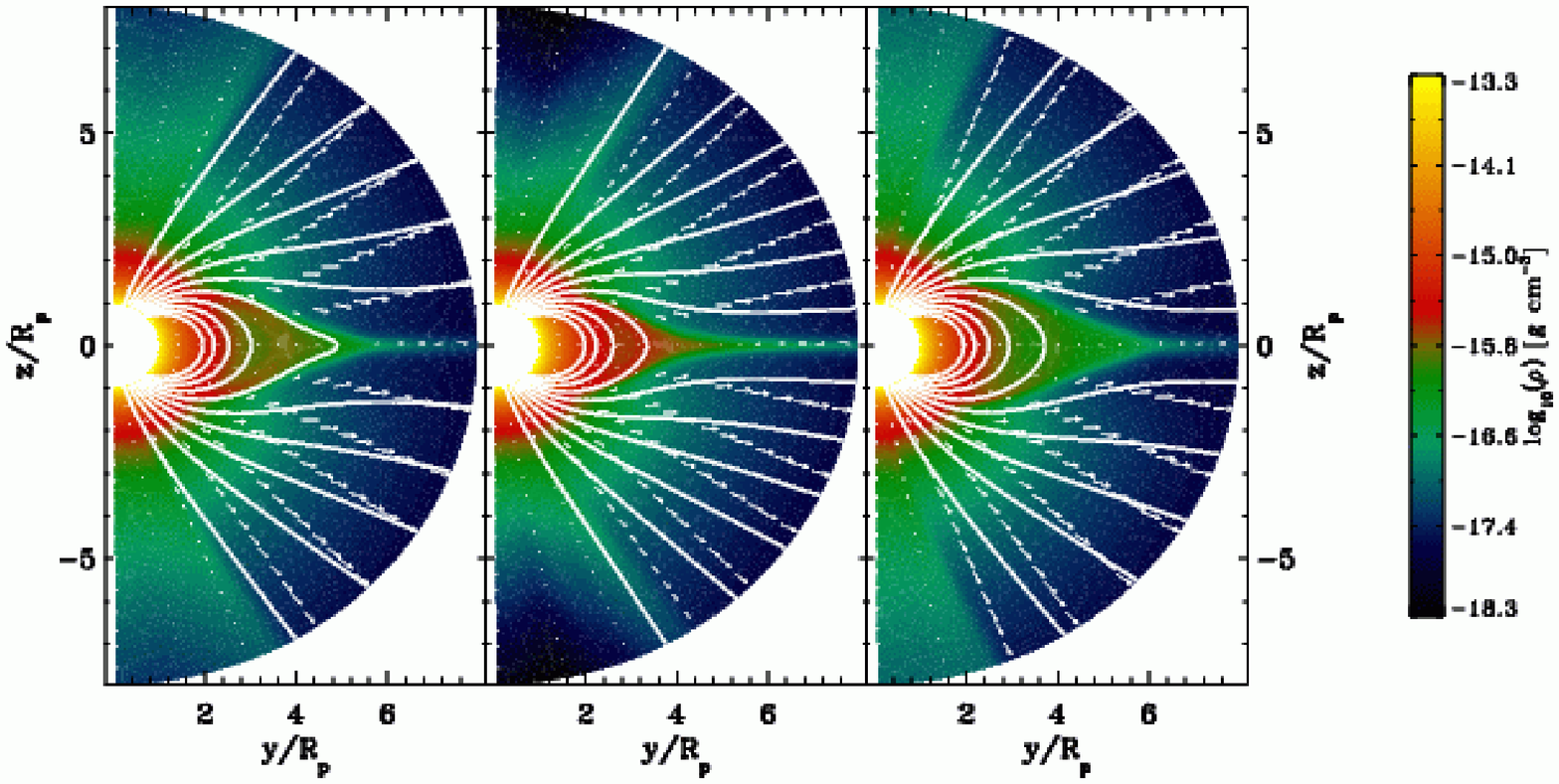}
\vspace{-8pt}
\caption{ 
Similar to Figure \ref{fiducial_rho.fig}, illustrating how the stellar
tide influences the size of the equatorial/polar DZs. All panels have
the same parameters as Model 1 of Table \ref{summary.tab}, except for
the varying orbital distance $D$. {\it Left}: the fiducial Model 1 for
HD 209458b with $D=0.047$ AU.  {\it Center}: Model 8 ($D=0.035$ AU),
which illustrates the effect of increasing the stellar tide by shrinking
the planet's orbit. {\it Right}: Model 9 ($D=0.06\ $AU). Note that the polar DZ
size is larger and equatorial DZ smaller for the stronger tide case (middle panel), as can be seen by the
range of angles with zero-length velocity vectors (white arrows).
As in Figure \ref{fiducial_rho.fig}, the innermost magnetic field lines
have been suppressed for clarity.
}
\label{fiducial_rho_tide.fig}
\end{figure*}

\begin{figure*}[t]
\epsscale{1.15}
\plotone{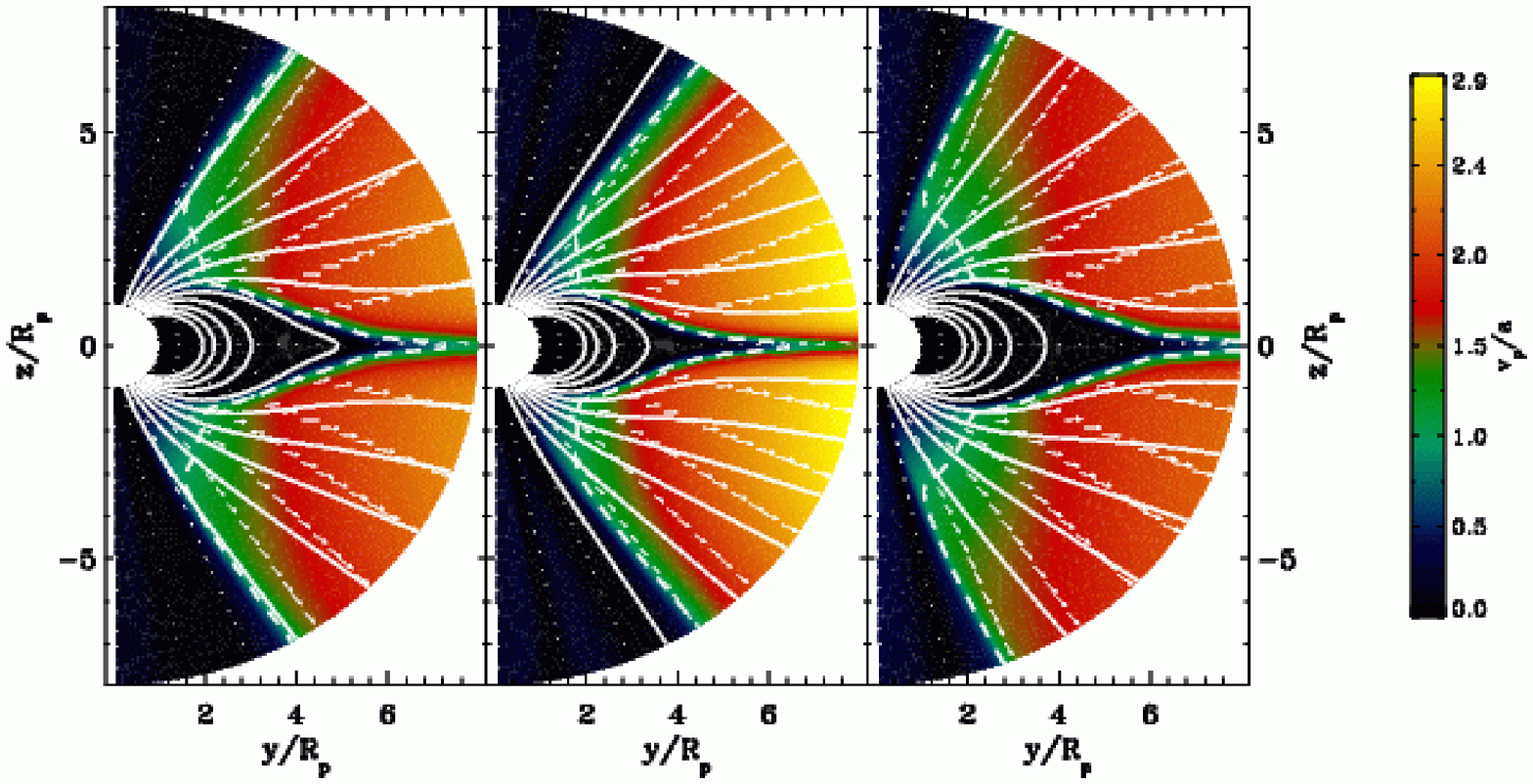}
\vspace{-8pt}
\caption{ 
Similar to Figure \ref{fiducial_rho_tide.fig}, showing the 2D structure of the
poloidal velocity $v_p$ field, in units of the isothermal speed $a$ for the same
three models shown in Figure \ref{fiducial_rho_tide.fig}. Higher stellar tide
pushes the slow magnetosonic point inward at mid-latitudes and leads to a
stronger outflow ram pressure, which can open a larger region of the planetary
magnetic field lines and decrease the size of the equatorial DZ. }
\label{fiducial_v1_tide.fig}
\end{figure*}

\subsection{Qualitative Results: Magnetic Field and Tidal Strength}
\label{sec:qualitative}

%
%
One of the most important qualitative results of this paper is the confirmation
of the three-zone structure of the magnetosphere predicted analytically in Paper
I. The three distinct regions are clearly visible in Figures
\ref{fiducial_rho.fig} and \ref{fiducial_v1.fig} --- (1) an equatorial dead-zone
(DZ) containing static gas confined by the magnetic field, (2) a wind-zone (WZ)
where an outflow is driven along open magnetic field lines, and (3) a second
polar DZ where the stellar tide has shut off the outflow (see also Fig.~7 of
Paper I). The range of magnetic field and stellar tide over which the equatorial
and polar DZ's exist has been discussed in Paper I. Roughly, the equatorial DZ
requires $\beta_0 \la 1$, i.e. the magnetic pressure dominates gas pressure at
the base of the atmosphere (the hydrogen ionization zone) at the equator. The
existence of the polar dead zone, and the inability to drive a transonic outflow
there, occurs inside a critical orbital separation. Roughly, this criterion
translates into the rotation velocity $\Omega_p$ at the fiducial sonic point
radius $r_{\rm s,0}=GM_p/3a^2$ must be supersonic, $\Omega_p r_{\rm s,0} \ga a$ (cf.~Equation 36 of Paper I).

The DZ-WZ boundaries in the simulation contain a shear layer separating the outflowing gas
in the WZ from the static gas in the DZ. In addition, the magnetic field changes rapidly in this 
boundary layer, implying a current sheet. The origin of this current sheet is that, for identical Bernoulli constant
at the inner boundary,  the WZ has 
smaller density compared to the neighboring DZ by a 
factor $\sim \exp(-v_p^2/a^2)$ (Mestel and Spruit 1987; Paper I), where $v_p$ is 
the poloidal wind speed. Since the total pressure, gas plus magnetic, must 
be continuous across the boundary, the decrease in gas pressure 
implies an increase in magnetic pressure, and hence a current sheet. Numerical issues related to the shear in velocity and magnetic field will be discussed further in the 
Appendix.

%
%
Figure~\ref{fiducial_rho.fig} illustrates the effect of the magnetic field 
on the density profile in the magnetosphere. The parameters for the 
runs in each panel are identical except for the magnetic field, with
$B_0=10,1$ and $50\ {\rm G}$ from left to right, respectively (Models 1-3 in 
Table~\ref{summary.tab}). As predicted in Paper I and expected
intuitively, the size of the equatorial DZ increases with the
field strength, when all other parameters are held fixed. For a dipole field line with a footpoint
at $\theta_0$, the magnetic pressure at the looptop at the equator decreases outward as
$B^2 \propto (1-3\sin^2\theta/4) \sin^{12}\theta_0 / \sin^{12}\theta$, and so field lines nearer the pole, with 
smaller $\theta_0$, suffer a larger decrease in magnetic pressure from pole to equator.
The larger DZ
size for larger $B_0$ then reflects the inability of ram pressure to overcome magnetic pressure, except
in a smaller region near the pole where the field decreases outward more rapidly.

The observational implication of the increase of DZ size with magnetic field is
that more of the circum-planetary material is expected to be confined within the
static dead zone, which should make this region easier to probe with transit
observations (see Section \ref{transit.sec}). A weaker magnetic field $B_0 \ll
1\ {\rm G}$ would not significantly confine the gas, and a larger range of
latitudes will participate in the outflow.

%
%
%

%
%
To more clearly differentiate the polar/equatorial DZs from the 
neighboring WZ, we plot in Figure \ref{fiducial_v1.fig} the poloidal
velocity for the same Models 1-3 
shown in Figure~\ref{fiducial_rho.fig}. The more dramatic contrast 
between the static DZ regions, where the fluid velocity is very 
subsonic, and the WZ with transonic outflow further illustrates 
the existence of the polar/equatorial DZs. The changing colors 
show the accelerating outflow in the mid-latitude regions along 
open magnetic field lines that have been combed out by currents 
in the magnetosphere. The darkest regions can be compared to the 
highest density regions in Figure \ref{fiducial_rho.fig}. 

%
%
Besides the field strength $B_0$, the structure of the magnetosphere 
is also influenced by the stellar tide. The tidal effects are 
illustrated in Figures \ref{fiducial_rho_tide.fig} 
and \ref{fiducial_v1_tide.fig}. All parameters except $D$ are held
fixed, even though temperature would likely increase as the planet is
moved nearer the star. The left panel is the fiducial 
Model~1. In the middle panel (Model 8), the orbital distance 
has been decreased to $D=0.035\ $AU, so that the stellar tide is
stronger than that for HD 209458b ($D=0.047\ $AU). As predicted in
Paper I, the stronger stellar tide increases the outward acceleration 
of the mid-latitude outflow by moving the sonic point inward. It 
results in an equatorial DZ that is slightly smaller in size but 
denser at the same distance from the planet relative to Model~1.  
Figure \ref{fiducial_v1_tide.fig} shows that the polar DZ size 
is also larger for the stronger tide case, as can be seen by the 
range of angles occupied by largely subsonic gas with small poloidal 
velocities, again in broad agreement with the analytic results of 
Paper~I.

%
%

\subsection{Quantitative Analysis: Density and Velocity Profiles}
\label{quantitative}

\begin{figure}[here]
\centering
\plotone{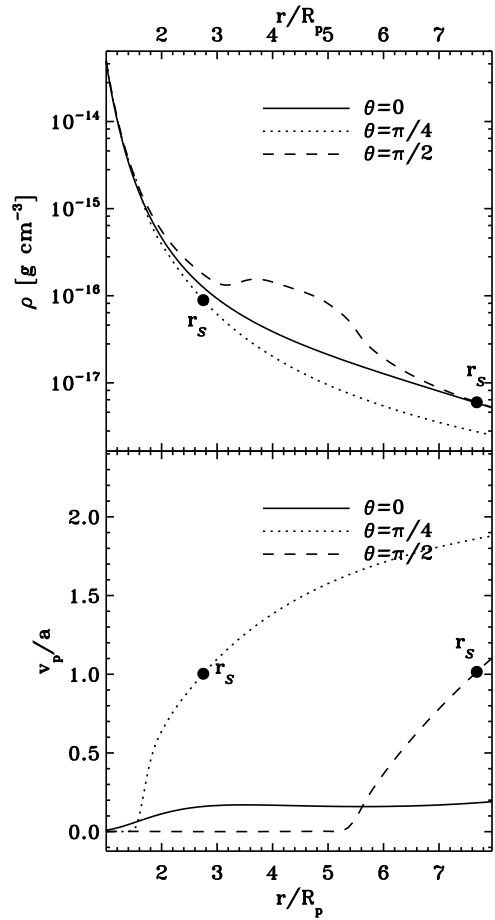}
\caption{
1D radial profiles of density $\rho(r)$ ({\it top}) and poloidal velocity
$v_p(r)$ ({\it bottom}) in units of the sound speed $a$ for the fiducial
Model 1 at three representative angles. The three
angles shown in each panel are $\theta \approx 0$ (solid black line),
$\theta \approx \pi/4$ (dotted black line), and $\theta \approx \pi/2$
(dashed black line). Sonic point locations are indicated by the black symbols labeled $r_{\rm S}$.
}
\label{fiducial_1D.fig}
\end{figure}

To examine the numerical simulations in more detail, 
Figure \ref{fiducial_1D.fig} shows the run of 
density and poloidal velocity along three different co-latitudes for the 
fiducial Model~1. The $\theta$ values are chosen to highlight the separate 
polar DZ, WZ and equatorial DZ regions, respectively. 

Along $\theta \approx 0$, near the pole, the density decreases 
rapidly with $r$ because the downward gravity of 
the planet and star must be balanced by pressure gradient in 
hydrostatic equilibrium. For this region, the flow speed remains 
well below the sound speed, in agreement with Paper~I, which 
predicts the absence of a transonic solution in the polar region. 
The $\theta\approx \pi/4$ line initially passes through the 
equatorial dead zone (where the poloidal velocity is close to 
zero; see the second panel of Figure \ref{fiducial_1D.fig}), 
before entering the wind zone. In the wind zone, the flow speed 
increases away from the planet, reaching nearly twice the sound 
speed at $r=8 R_p$. The density drops with distance accordingly. 

%
%
%

The density distribution along the equator at $\theta \approx \pi/2$ is the most
intriguing. After an initial rapid decrease, it increases for a short distance
near $r \simeq 3R_p$, and then resumes a slow decline. Such a ``bump" in the
density profile was predicted in Paper I, for the gas outside the Hill radius,
yet still confined inside the static magnetosphere. This is a consequence of the
outward pointing gravity outside the Hill radius in Equation \ref{eq:rl},
causing the density to increase outward instead of inward. However, the outward
increase in figure \ref{fiducial_1D.fig} occurs well inside $r_{\rm H} \simeq
5.4R_{\rm p}$! Hence it cannot be due to the change in the sign of gravity. In
the Appendix the origin of this density increase is explored, and seems to be
due to viscous stresses associated with numerical effects near the equatorial
DZ/WZ boundary. As the numerical resolution is increased, the density bump in
figure \ref{fiducial_1D.fig} decreases in size. In the Appendix it is shown that
increasing the resolution has a $\la 1\%$ effect on the integrated transit
depth, even as the density bump decreases. This gives confidence that
resolution-dependent effects are not leading to large errors in the transit
depth.


\subsection{Mass and Angular Momentum Loss Rates}
\label{rates}

The MHD simulations presented in this paper allow a more accurate determination of the rates of mass and angular momentum losses  ($\dot{M}$ 
and $\dot{J}$) as compared to the semi-analytic solutions from Paper I, since here the 
magnetic field geometry and fluid velocity are self-consistently computed.
These quantities are computed as a function of $r$ by integrals over $\theta$: \be
\dot{M}(r) &=& 
2\pi r^2 \int_0^\pi d\theta \sin\theta \rho(r,\theta) v_r(r,\theta)
\label{Mdot.eq}
\ee
and
\be
\dot{J}(r) &=& 
2\pi r^3 \int_0^\pi d\theta \sin^2\theta
\nonumber \\ & \times &
\left[ \rho(r,\theta) v_r(r,\theta) v_\phi(r,\theta)
 - \frac{B_r(r,\theta)B_\phi(r,\theta)}{4\pi} \right].
\label{Jdot.eq}
\ee
Typically $\dot{M}(r)$ and $\dot{J}(r)$ are constant with radius 
to better than 1\%,
which provides a check on the accuracy of the numerical
solutions. Table 
\ref{actualMJ.tab} summarizes the results for $\dot{M}$ and $\dot{J}$
for  Models 1-10.

The planet's magnetic field affects the dynamics in several ways.  
A stronger magnetic field increases
the size of the equatorial DZ,
restricting the WZ to a smaller range of latitudes. Therefore, one might
expect  that the mass-loss rate will decrease for a
stronger magnetic field.  This expectation is born out in the $\dot{M}$
values presented in Table \ref{actualMJ.tab}, where the mass loss decreases by
$\sim 35\%$ for a factor of 5 increase in $B_0$ from Model 1 to 
Model 3. Despite the reduction in $\dot{M}$, the total angular 
momentum loss rate $\dot{J}$ increased by a factor of $\sim  
2.5$, implying an increase in loss of specific angular momentum, 
$\dot{J}/\dot{M}$, due to a longer magnetic 
lever arm for the torque.
The effect of the magnetic field and tides on the specific angular momentum
loss is most clearly displayed in column 4
of Table \ref{actualMJ.tab}. The quantity $\dot{J}/(\dot{M}\Omega_{\rm p}R_{\rm p}^2)$
has the value $2/3$ ignoring these effects \citep{1968MNRAS.138..359M}, but is 
significantly larger here, even for relatively weak field cases.
Conversely, for a weaker magnetic field (i.e., Model 2), $\dot{M}$ 
is larger due to the larger range of latitudes in the WZ  (see the center panel of Figure \ref{fiducial_v1.fig}).

Stronger tides result in a slightly smaller equatorial DZ, because the
outward tidal force can open more magnetic field lines, but a larger 
polar DZ, due to the increased potential barrier.
Stronger tide also moves the sonic point inward, which tends to 
increase $\dot{M}$. For example, $\dot{M}$ of the
stronger tide Model 
8 is increased slightly, by a factor of $\simeq 20\%$, compared to
Model 1. Presumably if the tide is increased to the point that the sonic point
moves all the way in to the steeply-rising density profile deeper in the atmosphere, this
(figure \ref{fiducial_1D.fig}) will result in a greater sensitivity to the strength of the tide, as is expected for Roche
lobe overflow. 

A much larger change in $\dot{M}$ comes from varying
the base pressure $P_{ss}$ (Models 6 and 7) or the isothermal sound 
speed $a$ (Models 4 and 5). For example, when $P_{ss}$ increases by 
a factor of 10, from 0.05 to 0.5 $\mu$bar, $\dot{M}$ rises by a
factor of 16.9. When $a$ increases by $10\%$, from 10 to 11~km/s, 
$\dot{M}$ shoots up by a factor of 4.16! 
In the more heavily mass-loaded
winds, the field lines bend backward significantly in the 
azimuthal direction relatively close to the planet, forcing the 
fluid to rotate substantially below the corotation speed. The
self-consistent treatment of the deviation from corotation here is an 
improvement over the analytic solutions of Paper~I. Conversely, a
smaller $P_{ss}$ or $a$ leads to a lower $\dot{M}$, and a wind that 
is dominated by the magnetic field out to a larger distance. It is
interesting to note that the ratio ${\dot J}/({\dot M}\Omega_pR_p^2)$ 
has rather large values of 404.68 and 455.36 for Model 4 ($a=9$~km/s) 
and 6 ($P_{ss}=0.005$~$\mu$bar), respectively. They are very different
from the purely hydro winds from the planet, where the ratio is $2/3$. 
The relatively low mass loss rate in these cases allows the magnetic 
field to effectively enforce corotation up to a distance of $\sim 
20~R_p$.
%
%

The large spin-down torques found in the strongly magnetized models
may torque the planet away from synchronous rotation (Paper~I). 
Defining $\Gamma=\dot{J}/(\dot{M} \Omega_{\rm p}
R_{\rm p}^2)$ and $J = \alpha M R_{\rm p}^2 \Omega_{\rm p}$, the spindown
timescale is
\be
\frac{J}{\dot{J}} &  = &  \frac{\alpha}{\Gamma} \left( \frac{M}{\dot{M}} \right)
\nonumber \\ & \simeq &  3 \times 10^7\ {\rm yr}\ 
\left( \frac{\alpha}{0.1} \right) 
\left( \frac{660}{\Gamma} \right)
\left( \frac{M}{0.7M_{\rm J}} \right) 
\left( \frac{ 2\times 10^{11}\ {\rm g\ s^{-1}}  }{ \dot{M} } \right),
\label{eq:twind}
\ee
for Model 3 parameters. In torque equilibrium between magnetic spin-down torques and gravitational tidal torques, a steady-state asynchronous spin rate would occur, with associated steady-state gravitational tide heating. However, deviations from synchronous rotation depend on strength of the planet's tidal dissipation, which is uncertain, but likely to give synchronization timescales orders of magnitude shorter than Equation \ref{eq:twind} (e.g. \citealt{2003ApJ...589..605W}). For HD 209458b, the heating rate can be estimated to be far smaller than Jupiter's luminosity ($3\times 10^{24}\ {\rm erg}$) for gravitational tide synchronization timescales shorter than 1 Myr. Hence the magnetic spin-down torque and associated asynchronous rotation are not likely to give rise to a heating rate large enough to effect the thermal history of
the planet significantly.


\begin{deluxetable}{ccccc}
\tablewidth{0pt}
\tablecolumns{5}
\tablecaption{Mass/Ang. Mom. Loss Rates (HD 209458b)}
\tablehead{
\colhead{Run} &
\colhead{$\dot{M}$} &
\colhead{$\dot{J}$} &
\colhead{$\dot{J}/(\dot{M} \Omega_p R_p^2)$} &
\colhead{$\delta F/F$}
}
\startdata
Model 1  & 3.29  & 10.20 & 162.24 & 0.100 \\
Model 2  & 6.12  & 1.17  & 9.97   & 0.125 \\  
Model 3  & 2.11  & 26.40 & 655.64 & 0.157 \\   
Model 4  & 0.48  & 3.74  & 404.68 & 0.048 \\  
Model 5  & 13.70 & 19.05 & 72.81  & 0.209 \\
Model 6  & 0.25  & 2.20  & 455.36 & 0.028 \\
Model 7  & 55.50 & 37.59 & 35.46  & 0.470 \\
Model 8  & 3.89  & 15.24 & 131.92 & 0.154 \\
Model 9  & 3.21  & 7.04  & 165.70 & 0.088 \\
Model 10 & 2.87  & 33.40 & 610.59 & 0.253 \\
\hline
\enddata
\tablecomments{
Mass-loss rates $\dot{M}$ [$10^{11}$\
g s$^{-1}$] and angular momentum loss rates $\dot{J}$ [$10^{28}$\ g\
cm$^2$ s$^{-2}$] for the 9 models with parameters specified in Table \ref{summary.tab}, along
with the corresponding integrated Ly$\alpha$ transit depth $\delta F/F$
 (see Equation \ref{transit_integrated.eq}) from -200 to +200 km s$^{-1}$ from line center.
}
\label{actualMJ.tab}
\end{deluxetable}


\section{Transit Depths in Ly$\alpha$}
\label{transit.sec}

Section \ref{sims.sec} described numerical solutions for the MHD variables $\rho$ and $\vec{v}$ for different model parameters. In this section the mass density $\rho$ is converted into atomic hydrogen number density $n_{\rm H}$, and the transmission spectra for the models in Table \ref{summary.tab} are discussed.

As a point of departure when considering transmission spectra of the MHD simulation results, the simple model of \citet{2008A&A...481L..83L} is first summarized. They consider a plane parallel, isothermal atmosphere with base radius $R_{\rm b}$ and altitude $z=r-R_{\rm b}$. The number density is then $n(z) = n_0 \exp(-z/H)$, where $H=k_b T/(\mu m_p g)$ is the scale height, $\mu$ is the mean molecular weight, and $g=GM_{\rm p}/r^2$. The path length through the atmosphere is $\ell \simeq \sqrt{ 2\pi R_{\rm b} H}$, giving an optical depth $\tau_\nu(z) = n_0 \sigma_\nu \ell \exp(-z/H)$, where $\sigma_\nu$ is the Ly$\alpha$ (1s $\rightarrow$ 2p) cross
section. Setting $\tau_\nu(z_\nu)=1$ gives the altitude 
\be
z_\nu & \simeq & H \ln \left( \frac{1}{ n_0 \sigma_\nu \sqrt{ 2\pi R_{\rm b} H }  }\right) 
\ee
up to which the atmosphere is optically thick. The transit depth is then
\be
\frac{R^2_{\rm p}(\nu)}{R_\star^2} & \simeq & \frac{ R_{\rm b}^2 + 2R_{\rm b} z_\nu}{R_\star^2}.
\ee
The altitude $z_\nu \propto H \propto T/\mu g$, so hot atmospheres of low mean molecular weight gas around planets with low gravity will have large scale heights and transit depths.
Due to the steeply falling density, the transit depth has only a weak logarithmic dependence on $\sigma_\nu$. 

For the Ly$\alpha$ transit depths of the MHD models considered here, the DZ is hydrostatic, but the tidal/rotational forces are important, and so gravity is weaker than $GM_{\rm p}/r^2$. The corresponding larger scale heights make the plane parallel limit inaccurate, and the density profile, even of isothermal models, tends not to fall as steeply as it does deeper in the atmosphere. One consequence of the large scale heights is that there can be a significant contribution from gas with optical depth $\tau \leq 1$, but which occupies a large area. Hence, the ``opaque disk" concept -- that all absorption can be idealized as occurring inside the $\tau=1$ contour -- may no longer be accurate. Hence, for careful work numerical integrations are required. However, the analytic model gives useful intuition and is simple.

\subsection{ Details of the Calculation }

Stellar Ly$\alpha$ photons passing through the planet's atmosphere can be absorbed or scattered out of the line of sight to the observer, causing a decrease in flux.  In addition, the interstellar medium (ISM) can absorb/scatter the light, most prominently in the Doppler core of the line. The spectrum observed at Earth is the combination of these two effects. If the in-transit flux is $F_{\nu}$ and the out-of-transit flux is $F_{\nu}^{(0)}$, the fractional decrease in flux, the transit depth, is $(F^{(0)}_\nu-F_\nu)/F^{(0)}_\nu$.

The optical depth through the planet's atmosphere is given by
\be
\tau_{\nu}(y,z) &=& \int dx\ n_H(x,y,z)\ \sigma_\nu(x,y,z)
\label{tauyz.eq}
\ee
where $n_{\rm H}$ is the number density of the atomic hydrogen in the 1s state, $x$ specifies the direction along the line of sight to the star, $y$ and $z$
are the perpendicular coordinates on the sky. This line profile is taken to be a Voigt function (e.g. \citealt{1979rpa..book.....R}) evaluated using the isothermal temperature $T$, and bulk fluid velocity is included by transforming the photon frequency from the planet frame to the rest frame of the fluid.

The transit depth will be expressed in terms of a frequency dependent planet radius, $R_{\rm p}(\nu)$,
which is defined as the radius of an opaque disk that is required to produce the same transit depth as the integral over the model atmosphere:
\begin{eqnarray}
\frac{ F^{(0)}_\nu - F_\nu }{ F_{\nu}^{(0)} } & \equiv &  \frac{R_{\rm p}^2(\nu)}{R_\star^2} = \int_{\rm star} dy\ dz\ \left[1-e^{-\tau_{\nu}(y,z)}\right],
\label{transit.eq}
\end{eqnarray}
where corrections due to limb darkening have been ignored for simplicity. The fractional decrease in flux in Equation \ref{transit.eq} is independent of ISM absorption, and depends solely on the planetary atmosphere. The integration over y and z extends over the stellar disk, where star has radius $R_\star$.

The frequency-integrated transit depth for the models is calculated as
\be
\frac{\delta F}{F} &=& \frac{\int d\nu\ I_{\nu}^{(\star)} \left(  \frac{R_{\rm p}(\nu)}{R_\star} \right)^2 e^{-\tau_{\nu}^{\rm (ISM)}}}
                            {\int d\nu\ I_{\nu}^{(\star)}e^{-\tau_{\nu}^{\rm (ISM)}}}
\label{transit_integrated.eq}
\ee
where 
\be
I_\nu^{(*)} & = &
 \left[ 1 + \left| \frac{\Delta v}{67\ {\rm km\ s^{-1}}} \right|^3 \right]^{-1}.
\label{eq:sumer}
\ee 
is a fit to the shape of the Ly$\alpha$ intensity of the Sun under quiet solar conditions
\citep{1997ApJS..113..195F}. In units of velocity from line center at frequency $\nu_0, \Delta v = c(\nu-\nu_0)/\nu_0$.
The limits of integration in Equation \ref{transit_integrated.eq} are $- 200\ {\rm
km\ s}^{-1} \leq \Delta v \leq 200\ \rm km\ s^{-1}$ as in \citet{Ben-Jaffel 2008}.
The ISM optical depth $\tau_\nu^{\rm (ISM)}\ $ is computed using 
the Voigt line profile evaluated with a temperature $T_{\rm
ism}=8000\ {\rm K}$ and a neutral hydrogen column $N_{\rm H, ism} = 10^{18.4}\
{\rm cm^{-2}}$ \citep{Wood 2005}. The Ly$\alpha$ line is completely absorbed within $\Delta v \simeq \pm 50\ {\rm  km\ s^{-1}}$ from line center by the ISM.

The HI number density $n_H(x,y,z)$ is
computed by assuming a balance between optically-thin photoionization and radiative recombination (cf.~Paper I, Section 8),
\be
J_0 n_{\rm H} & = & \alpha_B n_{\rm e} n_{\rm p},
\label{eq:pieq}
\ee
where  $J_0 \approx (6\ {\rm hr})^{-1}(0.047\ {\rm AU}/D)^2$ is the ionization rate for a Solar EUV
spectrum (Paper I), and 
$\alpha_B(T) \simeq 2.6 \times 10^{-13}\ $ cm$^3$
s$^{-1}$ (10$^4$ K/$T$)$^{0.8}$ is the case B radiative recombination rate
\citep{Osterbrock 2006}.  Assuming charge neutrality, $n_e=n_p$, and setting
$\rho = m_p (n_{\rm p} + n_{\rm H})$, Equation \ref{eq:pieq} has the analytic solution
\be
n_H &=& \left[\frac{\sqrt{J_0/\alpha_B + 4\rho/m_p} - \sqrt{J_0/\alpha_B}}
		   {2}\right]^2
\label{nH.eq}
\ee
At a number density $n_{\rm eq} = J_0/\alpha_B$
the gas at density $\rho$ is 50\% ionized with $n_H = n_p$. For $n_{\rm H} \ga n_{\rm eq}$,
the gas is mostly neutral, and vice versa for $n_{\rm H} \la n_{\rm eq}$. The use of
a constant $J_0$ above simplifies the problem by requiring only the local
gas density $\rho$ to evaluate $n_H$. 

\subsection{ Results for HD 209458b}
\label{sec:209458b}
 
\begin{figure}[here]
\centering
\plotone{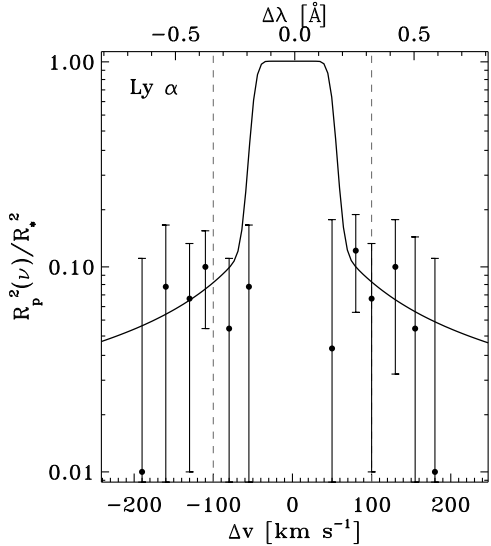}
\caption{Model 1 transit curve for comparison to the transit depths for HD 209458b from \citet{Ben-Jaffel 2008}.
}
\label{transit_M1.fig}
\end{figure}

Figure \ref{transit_M1.fig} compares the Ly$\alpha$ transit radius versus wavelength for the fiducial Model 1 to HST STIS data from \citet{Ben-Jaffel 2008}. Points near line center are heavily contaminated by ISM absorption and geocoronal emission and are omitted. Model 1 was designed to agree with the data through adjusting $P_{\rm ss}$ and $a$ (see Table \ref{actualMJ.tab}). The 
integrated transit depth, $\delta F/F \approx 10\%$ (see Table \ref{actualMJ.tab}), 
is in good agreement with \citet{Ben-Jaffel
2008} and \citet{Vidal 2008}. 

\begin{figure*}[t]
\epsscale{1.17}
\plottwo{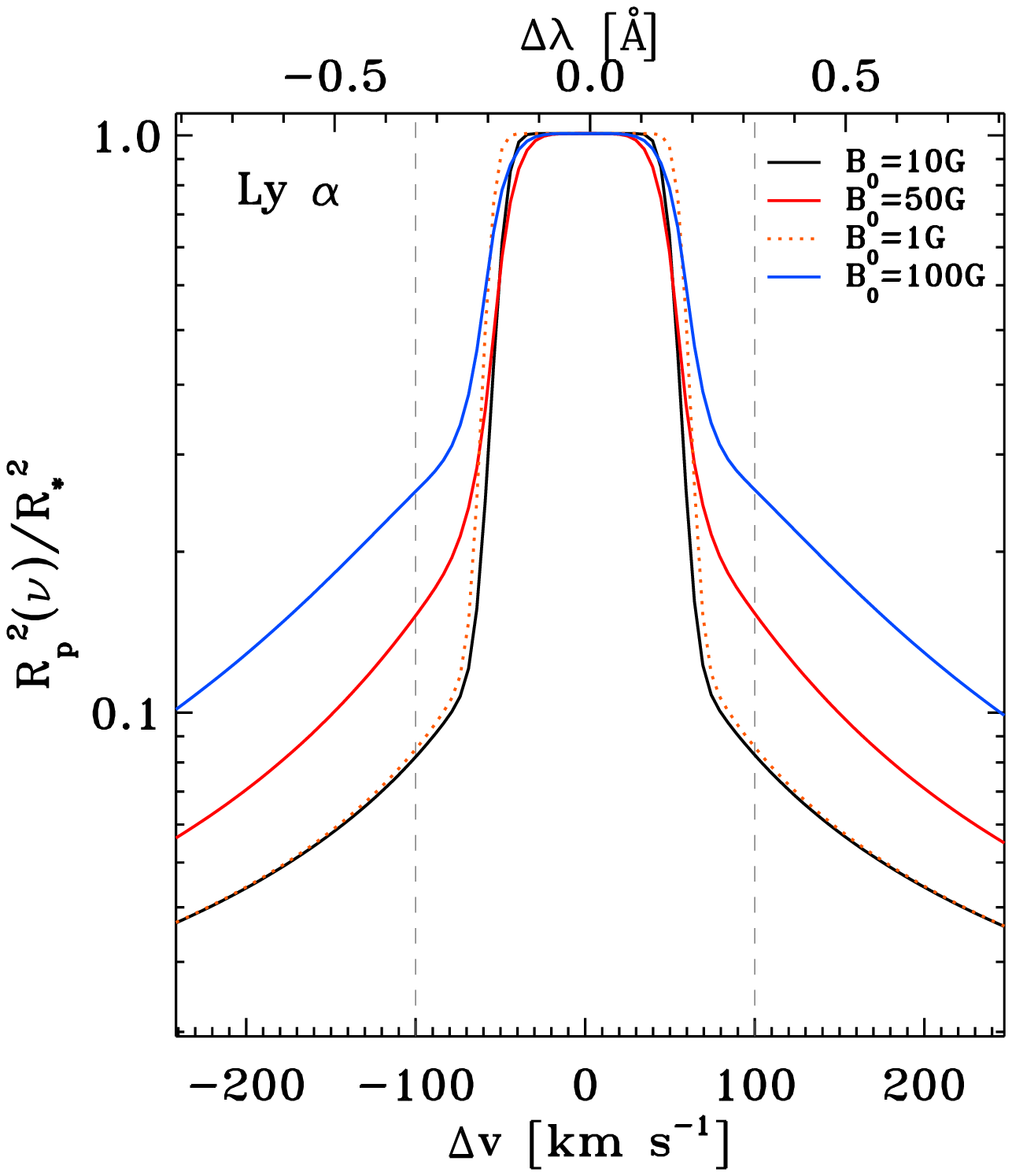}{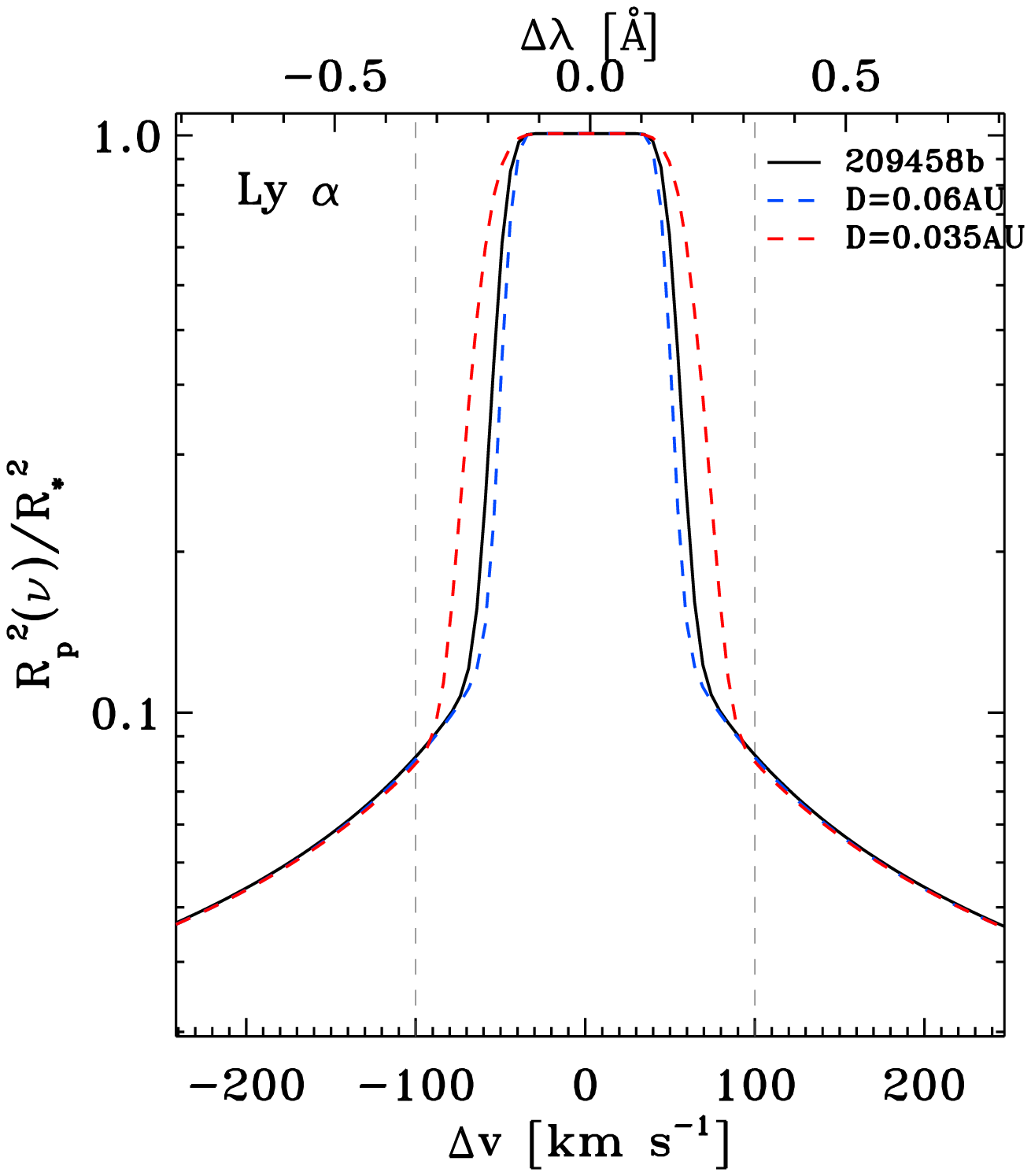}
\caption{ Ly$\alpha$ transit depth (Equation \ref{transit.eq})
versus wavelength in velocity units for selected models. The left
panel compares models with different $B_0$ (Model 1, $B_0=10\ {\rm
G}$, solid black line; Model 2, $B_0=1\ {\rm G}$, dotted orange
line; Model 3, $B_0=50\ {\rm G}$, solid red line; Model 10, $B_0=100\
{\rm G}$, solid blue line), while the right panel compares models
with different $D$ (Model 1, $D=0.047\ {\rm AU}$, solid black line;
Model 8, $D=0.035\ {\rm AU}$, dashed red line; Model 9, $D=0.06\
{\rm AU}$, dashed blue line). In each panel, only the one parameter
is changed, with all others held fixed at Model 1 values. For
clarity, bulk Doppler shifts are ignored in the left panel; they
only affect the line profile at $\Delta \vec{v} \la 50\ {\rm km\
s^{-1}}$ from line center. Bulk Doppler shifts are included in the
right panel, as the tidal force may give rise to additional
acceleration. Vertical lines are placed at $\Delta \vec{v} = \pm
100\ {\rm km\ s^{-1}}$ for reference.
}
\label{transit.fig}
\end{figure*}

Figure \ref{transit.fig} shows $R_{\rm p}(\nu)$ versus wavelength for
some of the models from Table \ref{summary.tab}. The left (right) panel shows the
effect of changing $B_0$ ($D$), holding all other parameters fixed. The model number
for each line is given in the figure caption.
For clarity, Doppler shifts $\Delta \nu = v_x(\nu-\nu_0)/c$ due to line of sight motion were
ignored in $\sigma_\nu$ in the left panel, but are included in the right panel, to assess the role of the tidal force in accelerating the fluid. Bulk fluid motion is able to increase the cross section significantly at wavelengths on the steeply falling part of the Doppler core, roughly within $\Delta v = \pm 50 $km/s of line center. 

First consider the effect of the magnetic field in the left panel of Figure \ref{transit.fig}. There is little difference between the $B_0=1$ and $10\ {\rm G}$ models, but in the range $B_0=10-100\ {\rm G}$, the transit depth is observed to grow on the wings of the line. Since bulk fluid velocity effects have been omitted, the increase in transit depth must be due to an increase in hydrogen column over a large area surrounding the planet.  Relative to $B_0=10\ {\rm G}$, there is an increase in $\delta F/F$ of 50\% for the $B_0=50\ {\rm G}$ model and 250\% for the $B_0=100\ {\rm G}$ model. This result clearly shows that the planetary magnetic field can have an important effect on the transit depth.

In the present paper, the base pressure and isothermal temperature
are parameters of the model, and the transit depth is most sensitive
to these two parameters. The range of these parameters (Table
\ref{summary.tab}) was based on the detailed one-dimensional
hydrostatic models, including ionization and heating/cooling balance,
presented in Paper 1. More complete MHD simulations including heating
and cooling would determine these quantities self-consistently as
part of the solution, and for a given stellar EUV heating rate they
would no longer be parameters. In such more complete models, the
magnetic field would still be an essentially unconstrained parameter,
as it is not measured or constrained by any observation as yet.
Figure \ref{transit.fig} shows that, if $B_0$ was the main uncertainty
in the model, an upper limit may be placed on magnetic field so
that the transit depth is not too large compared to observations.
For the fiducial parameters adopted for HD 209458b, that upper limit
would be $B_0\approx 50\ {\rm G}$. However, we caution the reader
that the large uncertainty in thermal structure due to uncertainty
in stellar EUV and accelerated particles fluxes likely limit the
practical ability to constrain the planetary magnetic field.
Nevertheless, for the parameters used in this paper, sufficiently
strong magnetic fields may in principle have a strong effect on the
transit depth.

Next consider the effect of changing the rotation rate and tidal force, by
changing $D$ with all other parameters held fixed. Comparison of the Model 1
lines (solid black line) in the left and right panels shows that Doppler shifts
due to bulk velocity in the WZ are small for the fiducial model and the model in
which the planet has been moved outward. However, moving the planet inward by
25\% to $D=0.035\ {\rm AU}$ has the effect of broadening the wavelength range
where $R_{\rm p}(\nu)$ is large (compare the dashed orange and solid black
lines). This is due to bulk fluid velocities Doppler shifting those wavelengths
to the Doppler core, where the cross section is large.

Gas that has escaped from the planet may still be strongly bound to the star,
and may achieve high bulk velocity due to the gravity of the star. In the
present case where the tidal force has been axisymmetrized, the effect is
symmetric on either side of the line. In the 3D case, red-shifted absorption due
to gas falling toward the star may achieve even larger velocities. For the
chosen box-sizes $r \sim 30R_{\rm p}$, the tidal force can accelerate fluid to
poloidal velocities $\simeq \Omega_p r \simeq 60\ {\rm km\ s^{-1}}$ in the
simulation box. For larger box sizes, even higher velocities may be achieved.
However, it is unclear from the present simulations if bulk velocities $\ga 100\
{\rm km\ s^{-1}}$ in the flow can affect the line profile, since the steeply
falling gas density may not be sufficiently large to give $\tau_\nu \ga 1$ at
such large distances from the planet. A further uncertainty is the interaction
with the stellar wind, which may confine the planetary wind to smaller radii,
with smaller acceleration by the tidal force.

\begin{figure}[here]
\centering
\plotone{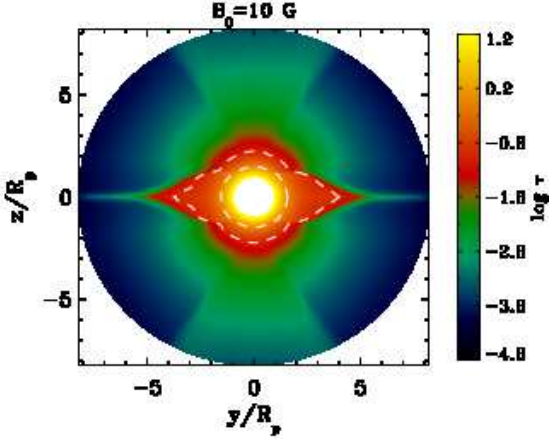}
\caption{ 
Contours of Ly$\alpha$ optical depth $\tau(y,z)$ at $\Delta v=100\ $km/s from 
line center for the steady-state simulation results from Model 1. The inner (outer) dashed
white line shows $\tau=1$ (0.1). 
}
\label{2D_tau_M1.fig} 
\end{figure}

\begin{figure}[here]
\centering
\plotone{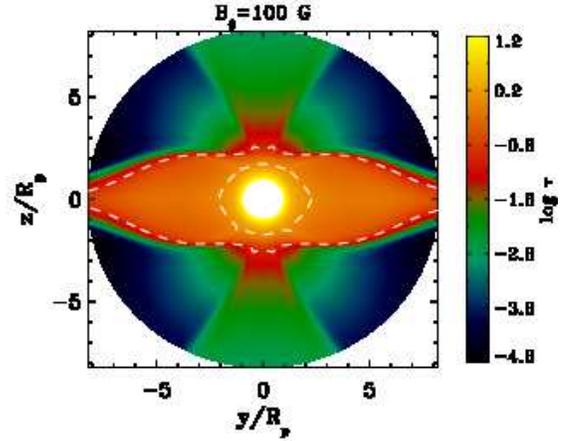}
\caption{ Same as Figure \ref{2D_tau_M1.fig} for the Model 10.}
 \label{2D_tau_M10.fig} 
\end{figure}

To understand the role of magnetic fields on the transit depths, contours of optical depth $\tau_\nu(y,z)$ at 100 km/s from line center are shown in the y-z plane in Figures 
\ref{2D_tau_M1.fig} and \ref{2D_tau_M10.fig}. The contours are evenly spaced in
$\log\tau$, and white dashed lines show the $\tau=0.1,1$ contours. The area
enclosed by the $\tau \ga 1$ contour is optically thick, and contributes
significantly to the transit depth. The region between $\tau=0.1-1$ contributes
to the transit depth proportional to $ {\rm area} \times \tau_\nu$, and so may
contribute significantly if the increase in area can overcome the decrease in
optical depth. For this to occur, the density must not decrease too rapidly
outward from the planet. Nearly the entire planetary upper atmosphere is
optically thick when observed near line center ($\Delta v$ = 0), but moving away
from line center the transit depth falls rapidly once the atmosphere becomes
optically thin, which occurs at a different value of $\Delta v$ for the range of
models shown.

Figure \ref{2D_tau_M1.fig} shows the steady-state result for Model 1 ($B_0=10\ {\rm G}$)
at an illustrative frequency corresponding to 100 km/s from the line center.
A large equatorial DZ with $\tau \sim 0.1-1$ extends out to $r=4R_{\rm p}$, while the same contour only extends to $r=2R_{\rm p}$ at the poles.
The gas in the mid-latitude WZ has significantly smaller optical depth compared to points in the neighboring polar and equatorial DZ's. The left-right asymmetry, most apparent in the $\tau=0.1$ contour, is due to the gas rotation Doppler shift in the cross section. Photons passing through the right side are shifted closer to line center, increasing the cross section, causing the contours to move further from the planet, and vice versa for the left side.

Figure \ref{2D_tau_M10.fig} shows the steady-state results for the stronger field Model 10 ($B_0=100\ {\rm G}$). The $\tau=0.1$ contour surrounding the equatorial DZ now extends to a significantly larger area which is sufficient to overcome the smaller optical depth there. 
Again, the mid-latitude WZ and polar DZ have far smaller optical depth compared to the equatorial DZ.
A comparison of Figures \ref{2D_tau_M1.fig} and \ref{2D_tau_M10.fig} clearly shows the growth of the optically thick equatorial DZ, which explains the increase in transit depth above $B_0 \simeq 10\ {\rm G}$ seen in the left panel of Figure \ref{transit.fig}.

Figures \ref{2D_tau_M1.fig} and \ref{2D_tau_M10.fig} clearly show the contribution to the transit depth from the DZ and WZ at a single, illustrative photon frequency ($100\ {\rm km\ s^{-1}}$ from line center). It is of interest to know what contribution the DZ and WZ make at all other wavelengths, and which wavelengths contribute most to the integrated transit depth in Equation \ref{transit_integrated.eq}. 
A technical point is that in order to know if a certain point is inside the DZ (WZ), one must trace along the field line to determine if it is closed (open), and if the fluid velocity is everywhere small (or accelerates to the sonic point). A simpler but approximate approach, followed here, is to compare the transit depth due to ``slow" and ``fast" material. The dividing line between the two is set by a threshold $v_{\rm p,thr}$ on the poloidal velocity; slow material has $v_{\rm p} \leq v_{\rm p, thr}$ and vice versa for fast material.  The slow material does not strictly trace out the DZ, since the fluid velocity at the base of the WZ is also small. By examination of optical depth contour plots using different velocity thresholds, we find that $v_{\rm p,thr}=0.1 a$ leads to only a small amount of slow material at the base of the WZ. Given the optical depths $\tau_{\rm slow, fast}(y,z)$ for the slow and fast material, the integrand of Equation \ref{transit_integrated.eq} can be computed. Note that while the optical depth is linear in the contribution from slow and fast material, the transit depth $\delta F/F$ is not, since $\tau$ occurs in an exponent.

\begin{figure}[here]
\centering
\plotone{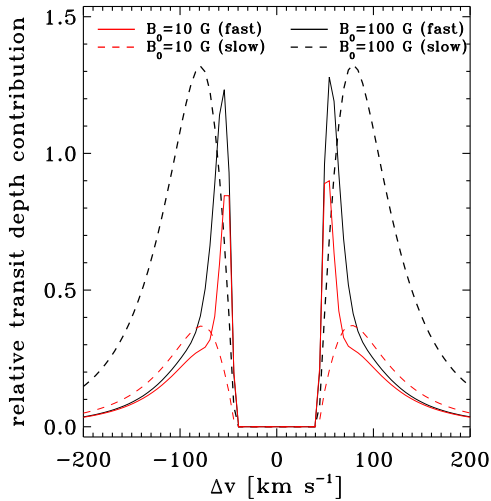}
\caption{ Integrand of the transit depth $\delta F/F$ in Equation \ref{transit_integrated.eq}, broken into the contribution from ``slow" fluid with $v_{\rm p} \leq 0.1a$ and ``fast" fluid with $v_{\rm p} > 0.1 a$. 
Two magnetic field strengths are shown: Model 1 ($B_0=10\ {\rm G}$) and Model 10 ($B_0=100\ {\rm G}$). }
 \label{dfi_M1L_vs_M1B100.fig} 
\end{figure}

Figure \ref{dfi_M1L_vs_M1B100.fig}  shows the integrand of Equation \ref{transit_integrated.eq}, separated into slow and fast material, and computed for two different field strengths, Model 1 (10 G) and Model 10 (100 G). First consider the $B_0=10\ {\rm G}$ lines. The contribution from the slow material shows the expected peak near $100\ {\rm km\ s^{-1}}$, with small contribution at large $\Delta v$ due to small stellar flux, as well as near line center, due to ISM absorption. The fast material shows narrow peaks just outside the region of ISM absorption. This is due to poloidal fluid motions Doppler shifting the photons from the Lorentzian wing back into the Doppler core of the line, where the cross-section increases rapidly toward line center. 
Even though the slow and fast materials make comparable contribution to the frequency-integrated transit depth (i.e., the areas under the dashed and solid curves are comparable), the shapes of the relative transit depth profiles are very different. This difference can in principle provide a way to distinguish the absorption due to slowly moving DZ material and fast moving WZ material. 
Next, comparing the $B_0=10$ and $100\ {\rm G}$ lines shows the far larger transit depth for the strong field case. This is due to the increased density at large radii, the which is the result of the larger dead zone extending out to a region where the gravity $-\grad U_{\rm rot}$ becomes quite small, so that the scale height is large and the density nearly constant. For the fast material, the higher field case shows absorption further from line center due to the higher poloidal velocity as the field is increased (see Figure \ref{fiducial_v1.fig}).

Having investigated the transit spectra in detail, the mass and angular momentum loss rates are now considered. Comparing the values of $\dot{M}$ and $\dot{J}$ to $\delta F/F$ in Table \ref{actualMJ.tab}, the dominant effect is that higher $P_{\rm ss}$ and $a$ lead to both larger $\dot{M}$ and $\dot{J}$, as well as $\delta F/F$. This is due to the higher gas density. The role of magnetic field well into the strongly magnetized regime is also clear, in that larger $B_0$ leads to larger $\delta F/F$ and $\dot{J}$, and slightly smaller $\dot{M}$.

How well did the semi-analytic solutions for $\rho$ and $\vec{v}$ in Paper I do at predicting the shape of the magnetospheres and the optical depth contours? Recall that in Paper I an inner dipole field was fitted to an outer monopole to represent the transition between wind and dead zones. The size of the dead zone was found by stress balance at the equatorial cusp point. Figure 14 of Paper I shows the hydrogen column for a set of models, and can be compared to the optical depth contours in Figures 
\ref{2D_tau_M1.fig} and \ref{2D_tau_M10.fig}. Model 1 (6) of Paper I is similar to Model 1 (3) here. The 3-zone structure is evident on both treatments, and many of the trends, e.g. the growth of the equatorial DZ, are evident in both. One key difference is that the equatorial DZ looks more ``cuspy" in the simulations where the field is self-consistently calculated, while the semi-analytic calculations give round-shaped dead zones. The dead zone sizes in the two calculations are comparable. The shapes of the wind zones differ between the two calculations due to the different shapes of the field lines, since the poloidal velocity is parallel to the poloidal field.
Overall, the semi-analytic methods of paper I give results in rough
quantitative agreement with the simulations here, at least near the planet where the assumed field shape of Paper I is approximately correct. Further from the planet, the backward bending of field lines, and the deviations from corotation become large, and are not taken into account in Paper I.

\subsection{ Results for HD 189733b }
\label{sec:189733b}

\begin{figure}[here]
\centering
\plotone{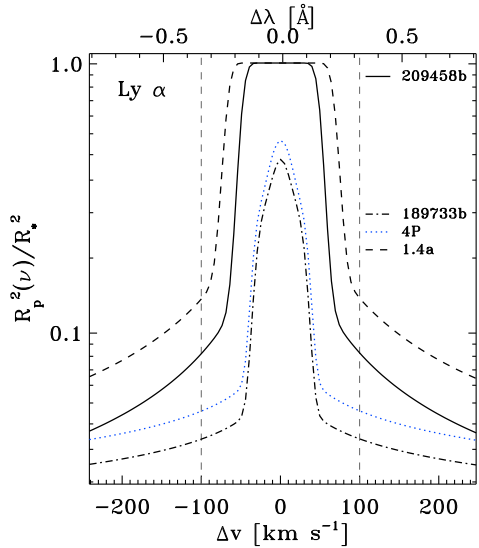}
\caption{The fiducial model for HD 209458b in comparison to three models for HD 189733b. The dot-dashed line is the frequency-dependent transit depth for parameters matched to HD 189733b, while the blue-dotted and dashed lines vary the base density and sound speed, respectively.
}
\label{transit_189.fig}
\end{figure}

\citet{Lecavelier 2010} used the HST ACS to measure an integrated transit
depth of 5\% in the Ly$\alpha$ line, smaller than the transit depth of
HD 209458b by a factor $\simeq 2$. In this section the exercise from
Section \ref{sec:209458b} is repeated for the planet HD 189733b, and
the effect of different $M_{\rm p}$, $R_{\rm p}$ and $D$ is discussed. The planetary mass and radius are set to $M_{\rm p}=1.14\ M_{\rm J}$ and $R_{\rm p}=1.14\ R_{\rm J}$, and the semi-major axis is $D= 0.031\ {\rm AU}$ (\url{http://exoplanet.eu/catalog}).  The Jean's parameter  $ \lambda = GM_{\rm p}/R_{\rm p}a^2$ is a factor $\simeq 2$ larger for HD 189733b than HD 209458b, for the same $a$. 

Figure \ref{transit_189.fig} compares the transit depth for the fiducial Model 1 of HD 209458b to three models for HD 189733b: a fiducial model with $P_{\rm ss}$, $a$ and $B_0$ identical to that of Model 1 of HD 209458b; a model with pressure 4 times larger; and a model with sound speed 40\% larger. The larger base pressure and sound speed are motivated by the higher stellar EUV flux for HD 189733b \citep{2011A&A...532A...6S}, due to the closer orbital separation, as well as higher stellar activity.

The fiducial model for HD 189733b has smaller transit depth ($\delta F/F=0.04$) than
that of Model 1 for HD 209458b ($\delta F/F=0.1$), due to the larger $M_p/R_p$ for HD 189733b causing
the density to decrease outward faster. The transit depth is surprisingly close to the observed value.
Since the EUV flux is higher, both the base pressure and temperature are expected to be higher than the HD 209458b case. As an illustration, increasing the base pressure slightly has the effect of increasing the density everywhere, leading
to larger transit depth ($\delta F/F=0.05$) comparable to the observed value for HD 189733b. Lastly, increasing
the sound speed by $40\%$ gives the model the same $\lambda$ as Model 1 of HD
209458b. The higher transit depth ($\delta F/F=0.24$), as compared to Model 1 of HD 209458b, is due, at least in part, to a stronger stellar tide (see Table~3).


\section{Summary}
\label{summary.sec} 

The MHD wind model simulations presented in this paper demonstrate that, for an
extended range of latitudes, the planet's magnetic field can qualitatively change
the properties of a
thermally-driven outflow. In Paper I, dipole field geometry and expected
field strengths of hot Jupiters were used to estimate the size of the
closed field line regions, where an outflow is quenched by rigid magnetic
field lines. The inclusion of the stellar tide was also shown to quench
the outflow in the polar region due to the higher potential barrier. The
estimated equatorial dead-zone sizes were $\sim 3-10R_{\rm p}$
for the parameters of interest.

MHD simulations
permitted the relaxation of a prescribed field
geometry, as hydrodynamic stresses and magnetic stresses in the
thermally-driven outflow from the hot inner boundary were computed
self-consistently through force balance both across and along
field lines. The simulation results verify the main features of the
semi-analytic models of Paper I,  including the dependence of their 
structure of the upper atmosphere on $B_0$ (Figures \ref{fiducial_rho.fig}
and \ref{fiducial_v1.fig}) and stellar tide strength (Figures
\ref{fiducial_rho_tide.fig} and \ref{fiducial_v1_tide.fig}). The MHD
wind model also permitted a self-consistent calculation of the mass
and angular momentum loss rates, which were presented in Section 3.3
for different stellar tide strengths and field strengths, 
and the integrated Ly$\alpha$ transit depth, $\delta F/F$
(the observable quantity, see Section 4). The results are most consistent with a
pressure of 50-60 nbar and a temperature of $\sim 10^4$ K at the base
of the thermosphere for HD 209458b, corresponding to a mass loss rate
$\dot{M} \simeq 3 \times 10^{11} \ {\rm g\ s^{-1}} $ and an angular momentum loss
rate $\dot{J} \simeq 6 \times 10^{28}\ {\rm  g\ cm^2\ s^{-2}}$.

A central result of this paper is that for sufficiently large
magnetic field, the large resulting equatorial DZ may dominate the
optically thick area which gives rise to the transit depth signal.
In this strongly magnetized regime, we find that the transit depth
increases with the magnetic field. If the thermal structure were
well known, this strong dependence on the magnetic field would allow
an upper limit to be placed on the planet's magnetic field so as
not to over-predict the transit depth. However, due to uncertainties
in heating rates due to stellar EUV and accelerated particles, such
an exercise is likely not possible. However, the parameter study in this paper
does make it clear what magnetic field strength is required to increase
the transit depth in the parametrized models, which may be compared to
the thermal structure in more realistic models.

A consequence of the existence of a large DZ in the strong field
  case is that 
Ly$\alpha$ absorption occurring out to near the Roche-lobe radius does
{\it not} directly imply the absorbing gas is escaping
(e.g. \citealt{Vidal 2003}), as emphasized in Paper I. The MHD model
(both analytic and numerical) does exhibit gas in the mid-latitude WZ which is escaping, however, the Ly$\alpha$ transmission spectrum is less sensitive to the gas in this region as the optical depths are lower (see Figure \ref{2D_tau_M10.fig}). 


Because of large observational uncertainties, the transit depth as a 
function of wavelength cannot precisely constrain the pressure at 
the base of the warm H layer (see Figure \ref{transit.fig}). The
integrated transit depths computed from the model Ly$\alpha$ spectra
presented in Section 4 provided another quantitative comparison with
observations. The high sensitivity of the integrated transit depth on the
pressure at the base of the warm H layer suggests that this observable
quantity can probe and constrain the conditions in the thermosphere of
highly irradiated hot Jupiters. At the same time, the numerical models
presented here provide complementary information about the resulting
expected mass and angular momentum loss rates, which are inaccessible
by observations.

\acknowledgements
The authors thank Duncan Christie and David Sing for helpful conversations. The authors also thank the referee for a thoughtful and thorough report which improved this paper. This work was supported in part by NSF (AST-0908079) and NASA Origins (NNX10AH29G) grants. 

\appendix
\section{Shear Layer and Current Sheet Near the Equatorial Dead-Zone/Wind-Zone Boundary}
\label{discussion.sec}

%
%
%
%

%
%

At the boundary between the equatorial DZ and mid-latitude WZ, there are 
sudden changes in fluid velocity and magnetic field over short distance. 
The origin of this shear layer and current sheet was discussed in Section
\ref{sec:qualitative}. As the simulations presented in this paper do not include
explicit viscous forces and Ohmic diffusion, it is the grid-scale numerical effects contained 
with the ZEUS-MP code that control the behavior of the solutions at these
discontinuities.

The possible effects from numerical diffusion came to our attention due to 
the spurious bump in density at the equator shown in Figure~\ref{fiducial_1D.fig},
where a rise in density occurs inside the Hill radius. This behavior contradicts
basic analytic considerations. It was shown in the 
Appendix of Paper~I that, in steady state, the Bernoulli constant
\be
W & \equiv & \frac{1}{2} v^2 + a^2 \ln \rho + U_{\rm rot}
\label{eq:Bernoulli}
\ee
(where $v$ and $U_{\rm rot}$ are defined in a frame corotating with the planet) 
must be a constant along a given field line in the dead-zone (see 
Equation~A10 of Paper~I). Since $v=0$, our choice of base density,
Equation~\ref{density}, indicates that $W$ should be a constant 
throughout the dead-zone. The force-balance equation~A8 of Paper~I 
implies immediately that the Lorentz force must vanish in the 
dead-zone. If this is the case, the density along the equator can
only increase with distance outside the Hill radius. This contradicts the
numerical results in Figure~\ref{fiducial_1D.fig}.

\begin{figure}[here]
\centering
\plotone{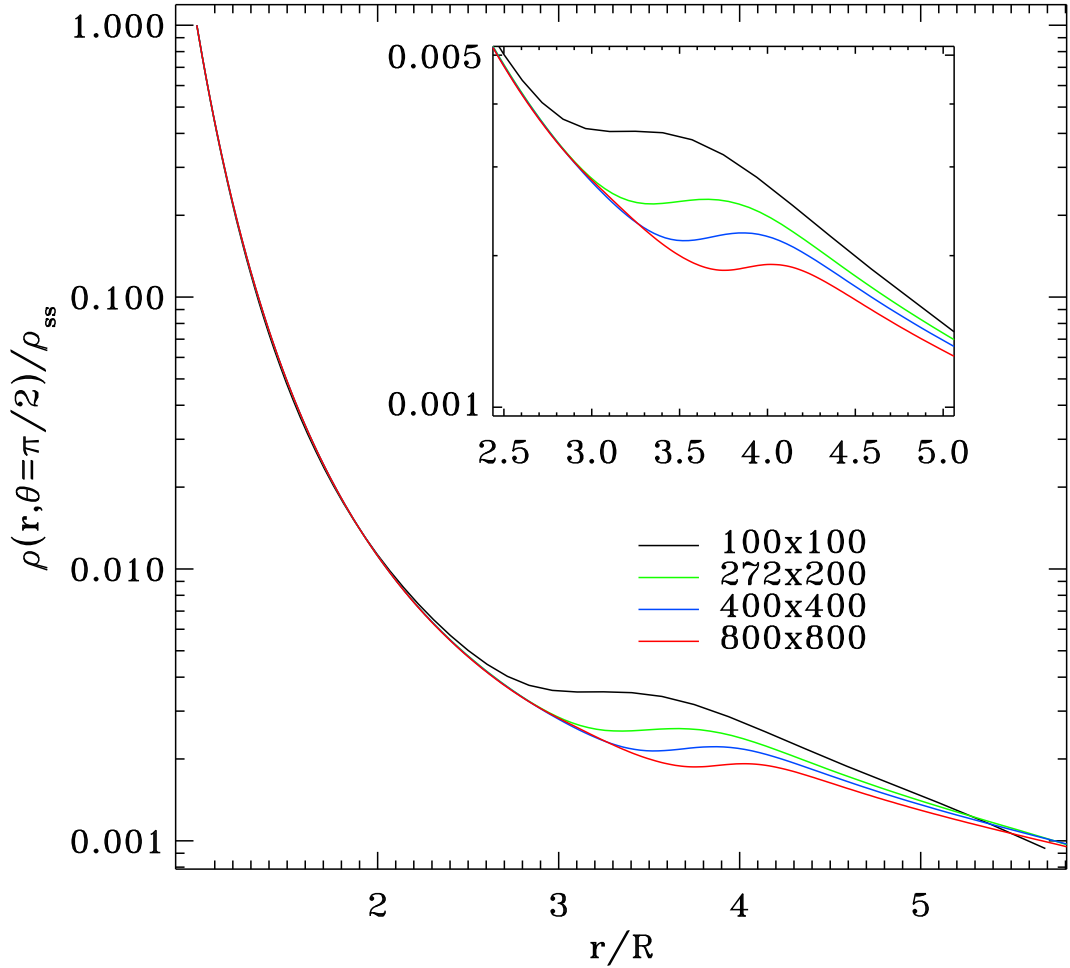}
\caption{
The equatorial density profile for Model 1 at three different grid resolutions
in steady-state, as well as an inset that shows a more detailed view of a region from $2.5 < r/R_{\rm p} < 5$.}
\label{dens_profiles_res.fig}
\end{figure}

The spurious density bump seems to be due to numerical effects near 
the dead-zone/wind-zone boundary, although the origin is rather subtle. 
In hindsight, it is not surprising
that the boundary would be difficult to treat numerically, because
there are discontinuities in all quantities: 
density and each of the three components of the velocity and magnetic 
field. The discontinuity in magnetic field, in particular, is
difficult to treat accurately (see, e.g., Fig.~16 of Stone \& 
Norman 1992b). One way to increase the accuracy is to increase the 
spatial resolution, which we have done for the fiducial Model~1. 
In Fig.~\ref{dens_profiles_res.fig}, we show the equatorial 
density profile at four different
resolutions ($100 \times 100$, $272\times 200$, $400\times 400$, and 
$800\times 800$), normalized by the substellar point density. 
It is clear that the spurious density enhancement decreases 
with increasing resolution, although even at $800\times 800$ it does
not disappear completely. 

\begin{figure}[here]
\centering
\plotone{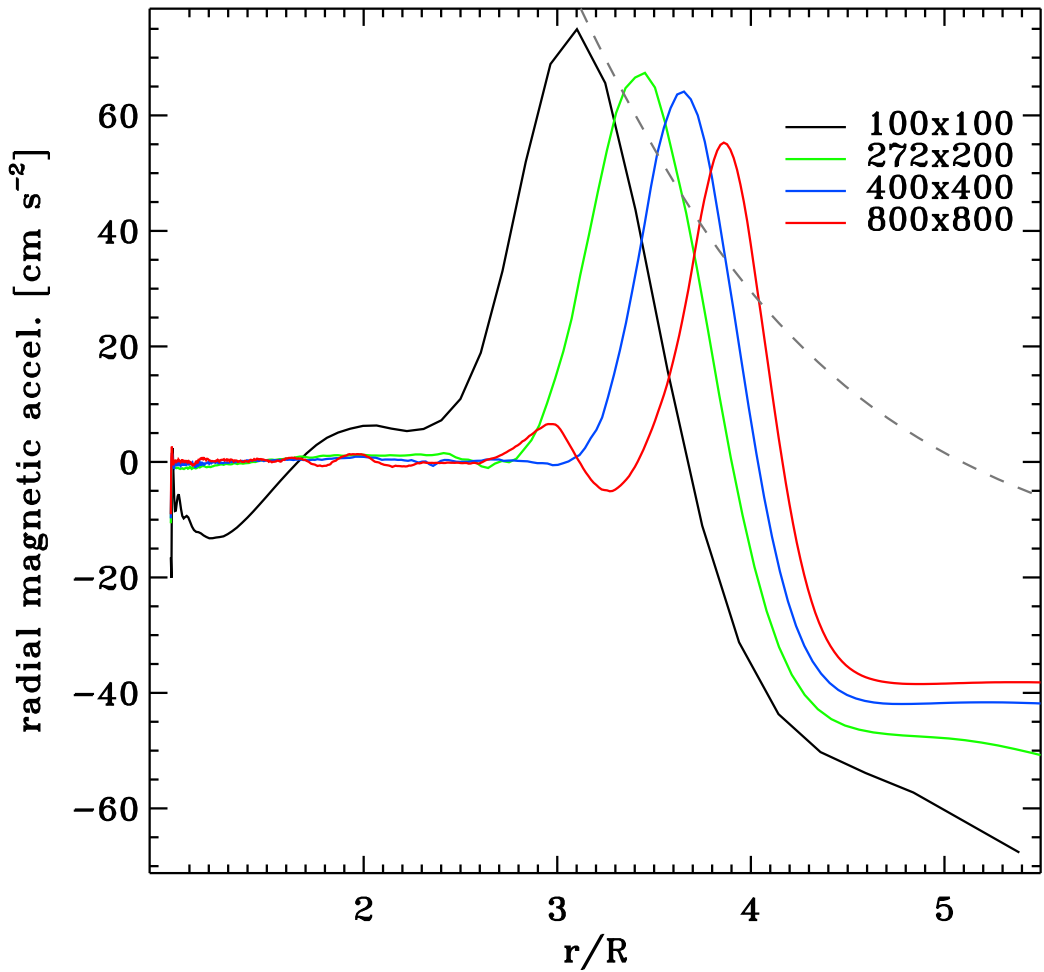}
\caption{
The radial magnetic forces along the equator for Model 1 at three different grid resolutions
in steady-state. The gray dashed line represents the effective gravity.
}
\label{forces_profiles_res.fig}
\end{figure}

\begin{figure}[here]
\centering
\plotone{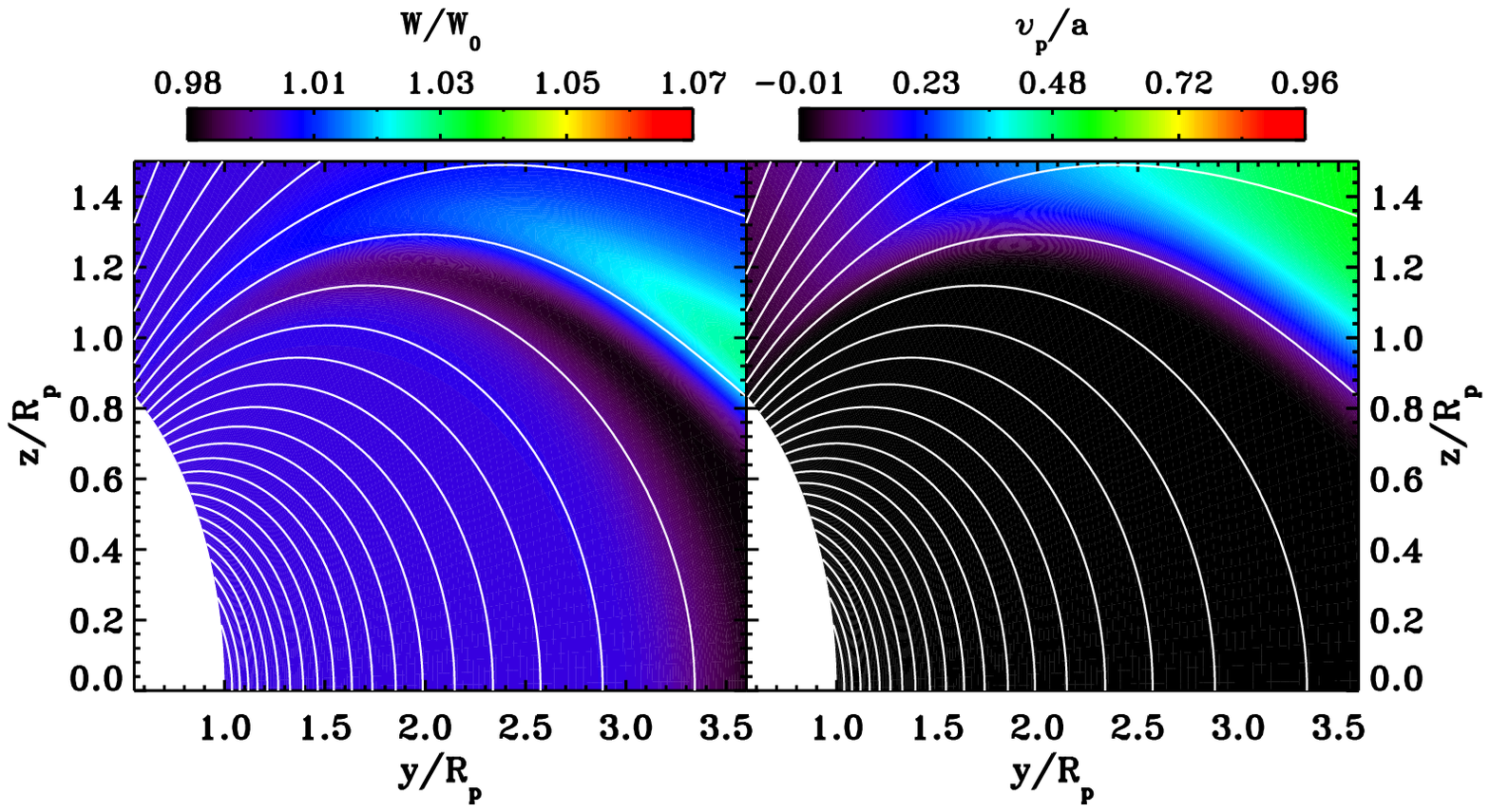}
\caption{
The variation of the Bernoulli constant relative to the base value for
a region including the DZ/WZ boundary layer, and the
poloidal velocity distribution for the same region. 
}
\label{W_vp_zoom.fig}
\end{figure}

This density enhancement is caused by magnetic forces (see 
Fig.~\ref{forces_profiles_res.fig}), which should vanish throughout 
the dead-zone according to the analytic considerations mentioned 
earlier. At the heart of these considerations is the constancy 
of the Bernoulli constant $W$ throughout the dead-zone. It  
breaks down in the numerical simulations, as illustrated in 
Figure~\ref{W_vp_zoom.fig}, where we show the distribution of 
the Bernoulli constant and poloidal velocity for a selected 
region for the standard resolution $272\times 200$. 
Note that $W$ is indeed very close to the expected value over 
most of the dead-zone (where the poloidal velocity is small, 
see the right panel), except in a layer near the DZ/WZ boundary, 
where deviation of order $1\%$ is evident; this is also the region 
where the magnetic forces become appreciable, and the density 
starts to increase outward spuriously. As the resolution increases, 
the boundary layer shrinks in size. 
This 
should serve as a cautionary tale for future simulations of hot 
Jupiter magnetospheres, especially in 3D, where the resolution 
will necessarily be coarser than in 2D. Nevertheless, the basic 
three-zone structure of the magnetosphere is robust. 

Although the density profile in the equatorial DZ is resolution dependent, 
the transit depth for the four models shown in Figure \ref{dens_profiles_res.fig}
varies only slightly, with values $\delta F/F=0.0943,0.1002, 0.1026,0.1031$
for resolutions 100x100, 272x200 (Model 1), 400x400 and 800x800. This variation
with resolution is far less than the variation from changing parameters in Models
1-9. It lends confidence that the
broad conclusions are not affected much by finite numerical resolution effects.

\end{document}